\pdfoutput=1

\RequirePackage{fix-cm}
\documentclass[twocolumn]{svjour3}          
\smartqed  
\usepackage{graphicx}
\usepackage{amssymb}
\usepackage{amsmath}
\usepackage{mathtools}
\usepackage{times}
\usepackage{bm}
\usepackage{natbib}
\usepackage{url}
\usepackage[switch]{lineno}

\usepackage{mathtools}

\begin{document}


\title{Global instability in the {Ghil-Sellers} model
}


\author{T. B\'odai \and V. Lucarini \and F. Lunkeit \and R. Boschi
}


\institute{T. B\'odai, V. Lucarini, F. Lunkeit, R. Boschi \at
              KlimaCampus, Meteorological Institute, \\ 
	      University of Hamburg, Hamburg, Germany \\
              \email{tamas.bodai@uni-hamburg.de}           
           \and
           V. Lucarini \at
              Department of Mathematics and Statistics, \\
              University of Reading, Reading, UK \\
              \email{valerio.lucarini@uni-hamburg.de}
}


\maketitle

\begin{abstract}
The Ghil-Sellers model, a diffusive one-dimensional energy balance model of Earth's climate, features -- for a considerable range of the parameter descriptive of the intensity of the incoming radiation -- two stable climate states (a warm one, taken to represent the present-day climate at the appropriate solar strength, and a cold one representative of snowball conditions), where the bistability results from the celebrated ice-albedo feedback. The unstable solution is obtained and characterized in this paper. We find such unstable states by applying for the first time in a geophysical context the so-called edge tracking method that has been used for studying multiple coexisting states in shear flows. This method has a great potential for studying the global instabilities in multistable systems, and for providing crucial information on the possibility of transitions when forcing is present. We examine robustness, efficiency, and accuracy properties of the edge tracking algorithm. We find that the procedure is the most efficient when taking a single bisection per cycle. Due to the strong diffusivity of the system trajectories of transient dynamics, initialized between the stable states with respect to the mean temperature, are confined to the heteroclininc trajectory, one which connects the fixed unstable and stable states, after relatively short transient times. This constraint dictates a functional relationship between observables. We characterize such a relationship between the global average temperature and a descriptor of nonequilibrium thermodynamics, the large scale temperature gradient between low and high latitudes. We find that a maximum of the temperature gradient is realized at the same value of the average temperature, about 270 K, largely independent of the strength of incoming solar radiation. Due to this maximum, a transient increase and nonmonotonic evolution of the temperature gradient is possible and not untypical. We also examine the structural properties of the system defined by bifurcation diagrams describing the equilibria depending on a system parameter of interest, here the solar strength. We construct new bifurcation diagrams in terms of quantities relevant for describing the thermodynamic disequilibrium, such as the temperature gradient and the material entropy production due to heat transport. We compare our results for the EBM to results for the intermediate complexity GCM PlaSim and find an interesting qualitative agreement. 
\end{abstract}

%

\section{Introduction}

In systems possessing multiple steady states, external perturbations can induce transition from one state to another. Near tipping points, small perturbations can cause large variations in the statistical properties of the system. The understanding of the properties of multistable systems and of the mechanisms behind the transitions between the various co-existing steady states is an emerging field of interest in many mathematical, natural science, and engineering contexts.

Recently many climate-related potentially multistable systems have been identified~\citep{Lenton12022008}. For example, a general circulation model (GCM) of an Earth-like model planet has been found, under e.g. present day solar forcing conditions, to feature bistability with a snowball cold state coexisting with a mostly snow-free warm state \citep{QJ:QJ543,Fraedrich:2012,BFSZ:2013,Boschi2013}. The physical mechanism responsible for such a snowball/snow-free bistability is the well-known ice-albedo positive feedback. This mechanism was singled out by 
\cite{TUS:TUS466} and 
\cite{Sellers:1969} when studying energy balance models (EBM). It has been suggested \citep{Hoffman28081998} that a snowball state is likely to have been realized by Earth's climate in the Neoproterozoic, when for millions of years data suggests a lack of biological activity in the ocean surface waters. 

A multi- or (for this discussion) bistable autonomous system has also at least one unstable state, which lies in-between the stable ones. This unstable state is embedded in the boundary of the basins of attraction to the two stable states. Initial states that fall on the basin boundary, which is a zero probability measure set, are attracted to the unstable state along its stable manifold. {The latter is in fact identical to the basin boundary. Being subject to smoothness properties of the governing equations, the unstable state is constituted most typically by a time-invariant {\em saddle set}~\citep{TG:2006}.}

Based on this argument, in all climate models where a bistability has been found there has to be an unstable state intermediate to the stable ones. Therefore, in the event of transition a perturbed orbit should closely visit the unstable state of the unperturbed autonomous system, and so it is expected that the unstable state leaves a fingerprint on the dynamics of the perturbed nonautonomous system. In particular, while small scale instabilities of the perturbed system are associated with the chaotic stable states of the autonomous system, the properties of the unstable state will be reflected in the large scale or global instabilities of the system.

We can distinguish between three types of transition scenarios in association with three types of forcing scenarios. One type of forcing is achieved by the slow change of a system parameter $\lambda$ in comparison with the characteristic time scales of the system. Starting from a parameter value that allows for bistability (e.g. the present day solar forcing considering Earth's climate), the slow change (decrease) of it gradually reduces the measure of the basin of attraction of the current (warm) stable state to zero. At such a critical value, $\lambda_c$, the system switches or tips abruptly to the other only remaining stable state. Hence, such a critical state of a system is referred to as a tipping point. Climate related tipping scenarios of so-called `tipping elements' are listed by \cite{Lenton12022008}. The central question in this context is if there is a way to predict the imminence of transition, possibly early on, and possibly even without having an accurate model of the system. The goal is to identify an effect in terms of an observable, called a precursor, which effect is universal to a large class of transition scenarios. {Candidates include the increased autocorrelation time \citep{Dakos23092008}, the increased variance \citep{GRL:GRL27358}, or the change of sign of the shape parameter of the extreme value distribution of some physical observable \citep{FLMW:2012}. The proximity of tipping may also be predicted by nonlinear softening \citep{Sieber13032012}.} The relevance of the unstable solution to tipping is established by the fact that at that point ($\lambda_c$) the branches of the bifurcation diagram belonging to the stable and unstable solutions `merge', as the basin boundary, in which the saddle set is embedded, shrinks onto the attractor. That is, the unstable state which is closely visited in the course of a transition is the one belonging to the critical parameter value $\lambda_c$. Such a transition has been referred to as B-tipping recently \citep{AWVC:2012}. The nonautonomous dynamics with a quasistatically changing forcing is essentially governed by the structural properties of the autonomous system, which is defined by the bifurcation diagram, i.e., the dependence of the equilibrium states on a chosen parameter, which also implies the sensitivity to a small change of that parameter.

A second forcing scenario is different from the first one in that the rate of change of the parameter can be comparable or faster than the unperturbed evolution. In larger than one degree-of-freedom systems this can bring about a transition for smaller changes of the parameter value $\lambda$ than that is needed with very slow forcing (approximately $\lambda_c-\lambda$). This is called a rate-induced transition, or R-tipping in short \citep{AWVC:2012}.

Third, processes that may be described by some stochastic or chaotic deterministic systems are seen as external perturbations when they are not the main process of interest. These perturbations have a fluctuating nature, in contrast with the first two types. The probability density and frequency spectrum are two basic descriptors of fluctuating processes. A powerful theory of stochastic dynamical systems exists for weak Gaussian white noise perturbations \citep{FW:1984}, by which e.g. the mean escape time from a basin can be calculated. A central element of the theory is a quasipotential, determined by the autonomous system, but it characterizes the steady state probability distribution of the perturbed system. It provides an intuitive picture insomuch that the unstable invariant set turns out to be situated on top of a potential barrier between the two attractors. More recently escapes from such a basin or potential well are called a noise-induced or N-tipping \citep{AWVC:2012}. However, even in the weak noise limit it is possible to have frequent transitions between two basins, so that the unstable set of the autonomous system is frequently visited. When this set is chaotic, the perturbed dynamics becomes chaotic, which is the much studied phenomenon of noise-induced chaos \citep{PhysRevLett.55.746,TLG:2008,LT:2011}. This situation can be generalized in two ways. First, one can consider stronger perturbations \citep{BKT:2011}, second, the perturbations can have complex variability (e.g. deterministic chaos versus red noise). Also in these more general situations it is expected that the unstable set influences the perturbed dynamics.

In summary, exploring the unstable solutions of a system may substantially contribute to the understanding of its general behavior since they are the key for large scale and global instabilities. In addition, not only the structural properties but also the transient dynamics related to the unstable solutions needs to be analyzed.

Numerical experiments with complex climate models have so far focused on the coexisting stable climates, {which are typically time-dependent -- presumably chaotic -- system states even without external forcing}, and on transitions between them {upon introducing some forcing} \citep{GRL:GRL22754,VM:2010,ISI:000291366800015,JGRD:JGRD11691}. For example, recent transition experiments with a GCM the Planet Simulator (PlaSim) \citep{QJ:QJ543,Boschi2013} indicated that thermodynamical properties \citep{PhysRevE.80.021118} are useful descriptors of the structural properties of the bistable system. The material entropy production, a quantity that measures how `far' the system is from thermodynamic equilibrium, can tell apart the warm and cold states much more firmly than the global thermodynamic properties like the average temperature. However, the analysis of unstable {climates that separate two stable climates has been limited to {\em time-independent} examples e.g. in a simple energy balance model \citep{Ghil:1976} and a more complex model of the 3D ocean circulation coupled to an EBM representing the atmosphere \citep{DW:2005}. In both of these examples the problem of numerically approximating the unstable solution in gridpoints reduces to finding a fixed point of a high-dimensional dynamical system by solving a system of algebraic equations. The latter analysis by \cite{DW:2005}, furthermore, involves a technique to determine the dependence of the solution on a system parameter, including situations when the solution changes stability at a certain parameter value \citep{Dijkstra:2005,Num_bif_meth}.} The reason for this {limitation in the analysis} is that so far no-one has attempted to search for {more realistic time-dependent} unstable climates in more complex models. 

\cite{Ghil:1976} studied a Sellers-type one-dimensional (1d) diffusive energy balance model, {which we refer to as the Ghil-Sellers model}. He solved the appropriate boundary value problem and found three coexisting climates, and thereupon carried out a stability analysis of these climate states. Ghil determined that an equilibrium state is stable/unstable when {the sign of a solution component of a certain initial value problem derived from the EBM is negative/positive, which sign} always corresponds to that of the first eigenvalue of the linearized evolution equation. {However, the eigenvalue itself, which gives the main time scale\footnote{{In our case the eigenvalue equals the escape rate, which latter is the reciprocal of the average life time in some neighborhood of the unstable state~\citep{LT:2011}.}} of the unstable transient process, is not determined by this procedure. The classical method of doing that, by solving a boundary value problem, is also detailed by Ghil.}

{We note that beside the heuristic argument above for the existence of an unstable state in a bistable system based on the concept of the basin boundary, there is a rigorous proof concerning autonomous dissipative `gradient systems', i.e., those in which time-independent forces have a potential: the mountain pass theorem \citep{Jabri:2003}. Asymptotic solutions of such systems, including infinite-dimensional ones, are time-independent. The existence of a potential in case of the Ghil-Sellers model was demonstrated by \cite{Ghil:1976} through providing a variational formulation of the problem. 
In autonomous systems where not all forces are potential, the steady states may be time-dependent. When such a system is bistable, the Freidlin-Wentzell (\citeyear{FW:1984}) theory defines a quasipotential, with which the mountain pass theorem can still be applied. In case of {\em non}autonomous bistable systems we can retain our heuristic argument. However, the unstable saddle set and its stable manifold (the basin boundary) is not time-invariant, and it can be defined in a {\em pullback} sense \citep{PhysRevE.87.042902}.} 

In more complex models like GCMs the unstable states can be expected to be high-dimensional chaotic sets, implying {\em time-dependent} steady states. In a finite window of time the evolution of the unstable state can be posed, in principle, as the solution of a boundary value problem, for which the initial and final states are prescribed (besides spatial boundary conditions). In order for a boundary value problem solver algorithm to converge to a solution, it is necessary that the prescribed initial and final states belong to the {time-dependent} unstable set. Providing such states is not a trivial task; {and we believe that in general it is not even possible}. To track down the possibly complicated unstable climates in various models, we intend to use a very different approach. We bracket the boundary of the two basins by two numerical trajectories, one in each basin, which are reinitialized to closely bracket the boundary again once diverged from it (and each other) to an unsatisfactory degree. This approach, called `edge tracking,' has been proposed and successfully applied by Eckhardt and coworkers \citep{PhysRevLett.96.174101,PhysRevE.78.037301,Schneider13022009} to track unstable solutions, or `edge states', of shear flow problems, such as pipe flow, plane-Couette flow, and plane-Poisulle flow, where the stable laminar flow can coexist with chaotic -- and depending on the problem -- stable or transiently turbulent flows. The edge tracking technique, conveniently, in contrast with the above mentioned alternative approach by solving a boundary value problem, does not require guess values on the unstable set, the edge state, but it can be initialized by the relatively easily obtainable stable steady states.

Before applying edge tracking to a complex climate model, the present study illustrates the methodology and its potential value utilizing the 1d Ghil-Sellers EBM. This model has been extensively studied before \citep{Ghil:1976,ROG:ROG799} and meets the necessary requirement by exhibiting bistability. We note that, in principle, the edge tracking algorithm may be applied to a zero-dimensional (0d) EBM as a `proof of concept' in the context of geophysical phenomena. However, the unstable equilibrium in that case is an overly simple object, a fixed point in a one-dimensional phase space, at the peak of a potential barrier, and so the application of the new method would yield rather limited insight. 
Furthermore, insightful thermodynamic nonequilibrium properties are not resolved by the 0d EBM either. In the direction of a higher-dimensional phase space where more complicated dynamical behavior is possible and thermodynamic disequilibrium is also represented, we favor a 1d EBM.

In this paper we also present a characterization of the structural properties of the {Ghil-Sellers} 1d EBM in the full range of bistability varying the solar strength, complete with the unstable branch of the bifurcation diagrams, in terms of thermodynamic quantities. Besides the average temperature, which is the subject of classical studies, we also consider quantities in association with thermodynamic disequilibrium, such as e.g. meridional heat transport, temperature gradients, or the material entropy production. We also characterize the transition from the unstable state to the stable ones resulting from infinitesimal perturbations of the unstable state, again in the full range of bistability, and again in terms of both the average and the nonequilibrium thermodynamics. This transition is governed by a heteroclinic orbit of the system, and it is associated with a constitutive relationship between the average and nonequilibrium thermodynamical properties. This relationship is found to depend only slightly on the solar strength, and e.g. the temperature contrast achieves a maximum at 270 K, when approximately half of the globe is snow-covered. 

The paper is organized as follows: Section \ref{sec:model} summarizes the main characteristics of the 1d Ghil-Sellers EBM utilized for our study. Section \ref{sec:methodology} introduces the edge tracking algorithm to explore the unstable solutions. In Sec. \ref{sec:results} results concerning the structural properties and the transient dynamics related to the unstable solutions are presented. In Sec. \ref{sec:conclusions} we close the paper with discussion and an outlook.

\section{{The Ghil-Sellers} one-dimensional energy balance climate model}\label{sec:model}

We adopt the Sellers-type one-dimensional energy balance model studied by 
\cite{Ghil:1976}. The only resolved spatial variable is the latitude $\phi\in[-\pi/2,\pi/2]$. The major processes considered by the zonally symmetric model are: radiation and meridional transport of heat. The latter process, facilitated by cyclonic eddies, is modeled in a crude way as a diffusive process, where a diffusion coefficient parametrizes the unresolved fluid dynamics. The model equation is essentially a forced heat equation for the {zonal-average `air'} temperature $T$ {extrapolated to the sea level}, written in a spherical coordinate system, for which a new variable $x=2\phi/\pi\in[-1,1]$ is introduced:

\begin{equation}\label{eq:pde}
  \begin{split} 
  c(x)T_t &= \left(\frac{2}{\pi}\right)^2\frac{1}{\cos(\pi x/2)}[\cos(\pi x/2)k(x,T)T_x]_x \\
  &+ \mu Q(x)[1 - \alpha(x,T)] \\
  &- \sigma T^4[1 - m\tanh(c_3T^6)],  
  \end{split}
\end{equation} 
with boundary and initial conditions:

\begin{equation}\label{eq:BC_n_IC}
  T_x(-1,t) = T_x(1,t) = 0, \quad T(x,0) = T_0(x).
\end{equation} 
The subscript e.g. in $T_x$ designates differentiation with respect to the variable $x$. The left hand side of Eq. (\ref{eq:pde}) represents the tendency, i.e., the rate of change, of the zonally averaged energy with $c(x)$ being the effective heat capacity of the atmosphere, land, and ocean per unit surface area at $x$. The model is symmetric to the Equator.

The first term on the right hand side (RHS) represents the meridional heat transport which is modeled by a Fourier-like law, so that it is proportional to the gradient $T_x$. Parameter $k$ is a combined diffusion coefficient depending on $x$ and $T$:

  \begin{align}
    k(x,T) &= k_1(x) + k_2(x)g(T),  \textrm{with} \label{eq:diff_coeff}\\
    g(T) &= c_4/T^2\exp(-c_5/T). \label{eq:latent_diff_coeff} 
  \end{align}
%
Here, $k_1$ and $k_2$ are eddy diffusivities for sensible and latent heat, respectively; {and the form of $g(T)$ is empirically constructed on the basis of 
the thermodynamics of moist air~\citep{BBB:1945}.}

The model is forced by heat absorption due to short-wave solar irradiation (second term on the RHS) and by the heat loss to space by long-wave emission (third term). The solar forcing is controlled by the solar irradiance $Q$ and the albedo $\alpha$. The long-wave radiation is represented by Boltzmann's law, modulated by the green house effect being represented by a plausible decreasing function of the temperature (in square brackets).

We can modulate the incoming solar radiation by changing the parameter $\mu$. The present day conditions are realized when $\mu=1$.

The following formulation of the latitude- and temperature-dependence of the albedo, including upper and lower cutoffs, defines the intensity of the ice-albedo feedback and is compatible with bistability:

\begin{equation}\label{eq:albedo}
    \alpha(x,T) = \{b(x) - c_1[T_m + \min(T - c_2z(x) - T_m,0)]\}_c, 
\end{equation}
where the subscript $\{\cdot\}_c$ means a cutoff, described by a generic quantity $h$ as follows:

\begin{equation}\label{eq:cutoff}
    h_c = \left\{
      \begin{tabular}{llcl}
	$h_{min}$, &                  & $h$ & $\leq$ $h_{min}$ \\
	$h$,       & $h_{min}$ $<$    & $h$ & $<$    $h_{max}$ \\
	$h_{max}$, & $h_{max}$ $\leq$ & $h$ &  
      \end{tabular} 
      \right.
\end{equation}
In Eq. (\ref{eq:albedo}) $c_2z(x)$ gives the difference between sea-level and zonally-averaged ground-level temperatures, and $b(x)$ is a temperature-independent component of the zonally-averaged albedo, describing the variation of the albedo with respect to the latitude only when the ground temperature exceeds $T_m$, in which case the ground is completely snow free.

Further details on the origin and interpretation of terms of the model can be found in \citep{Ghil:1976} and references therein. In particular, \cite{DP:1973} give an outline of a systematic derivation of the model. Empirical data {for one hemisphere} is adapted from \citep{Ghil:1976}, such as values of the empirical functions, $c(x)$, $Q(x)$, $b(x)$, $z(x)$, $k_1(x)$, $k_2(x)$, at discrete latitudes, and empirical constants, $c_1$,..., $c_3$, $\sigma$, $m$, $T_m$. {A numerical code written in MATLAB is publicly available on Mathworks' File Exchange website\footnote{{http://www.mathworks.co.uk/matlabcentral/fileexchange/46391-gsebm}}. (This code should be straightforwardly `portable', as a manual effort, to freewares such as Octave or Scilab.)} Note that we use a new value $b(\phi=85^{\circ})=2.912$ as a correction for a likely typographical error in \citep{Ghil:1976}. We found agreement with Ghil's solutions selecting this value, whereas the value reported in the paper leads to somewhat different temperature profiles. We also omit a redundant $10^3$ factor of $c_4$ given in the same paper.

With the original $\alpha_{max}=0.85$ we have been able to reproduce, very satisfactorily to a visual inspection, e.g. the temperature profiles shown by Fig. 3(a) of \citep{Ghil:1976} and the bifurcation diagram shown by Fig. 10.6 of \citep{GC:1987}. We note that to be able to reproduce the bifurcation diagram featuring the complete range of bistability, it was necessary to eliminate negative values of $k_2$, because with them $k$ itself becomes negative for about $\mu>1.03$, which violates the second law of Thermodynamics. We used the following new values before inter- and extrapolation to the computational gridpoints: $k_2(\phi=15)=2\cdot10^{-3}$ and $k_2(\phi=5)=10^{-3}$. We have checked that for $\mu<1.03$ (when the original negative $k_2$ values do not make the numerics unstable) the original and modified models yield very closely matching results. In the present study, however, we apply $\alpha_{max}=0.6$, with which the range of bistability is greatly reduced and the temperatures in the snowball conditions are substantially higher, yielding closer correspondence to the more realistic results produced by PlaSim shown by Fig. 1 of \citep{QJ:QJ543}. Also in this case we applied the new values for $k_2$ specified above.

\begin{figure}
\centerline{\includegraphics[width=\linewidth]{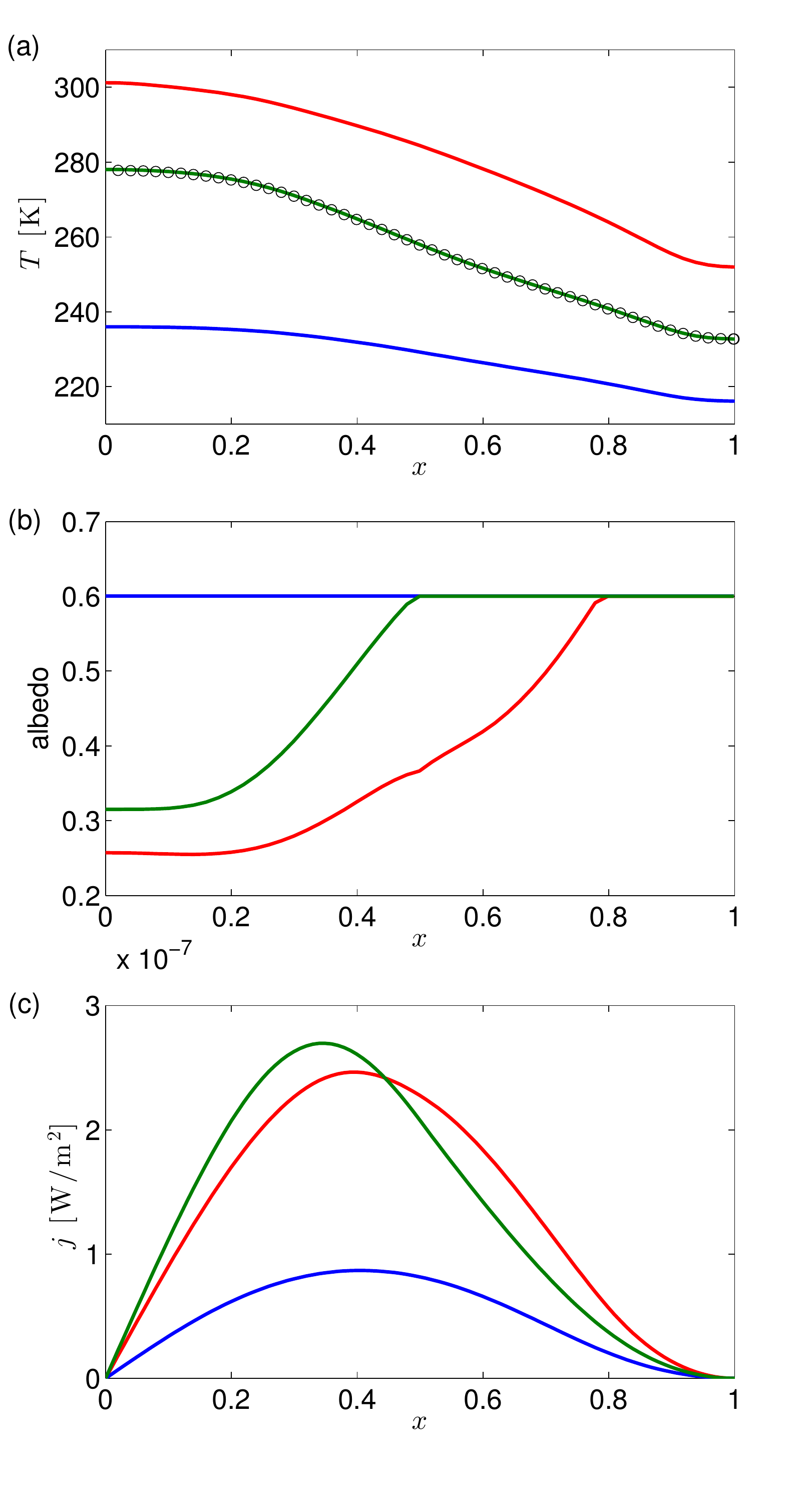}}
\caption{Temperature versus latitude profiles (a), being the stationary solutions of Eq. (\ref{eq:pde}), and corresponding (color-matching) albedo- (b) {and (specific) heat transport rate profiles (c)}. The unstable profile between the two stable ones, represented by hollow circle markers, was determined by the edge tracking procedure to be described in detail in Sec. \ref{sec:edgetracking}. The green solid line shows a reference numerical solution obtained by a boundary value problem solver, Matlab's \texttt{bvp4c}.} \label{fig:temp_profiles}
\end{figure}

Similarly to the original model studied by Ghil, for $\mu=1$ the modified model too has three equilibrium solutions in the range of physical interest, say $200<\lim\limits_{t \to \infty} T(x,t)<310$, two of which are stable, $T_W(x)$ and $T_C(x)$, and an intermediate solution being unstable, $T_U(x)$. These are shown in Fig. \ref{fig:temp_profiles}, accompanied by corresponding albedo {and heat `flux' or (specific) heat transport rate profiles. The latter is defined\footnote{{Equation (\ref{eq:pde}) can be rewritten concisely in the form $cT_t = -j_{\phi}/\cos(\phi) + F$, with $F$ comprising the radiative forcing terms; and $j=J/(A/2)$, where $J=H2\pi\cos\phi Rq$ is the integrated meridional heat transport in [Watts] at some latitude, and $A,R$ are respectively the surface and the radius of the globe. $J$ is a surface integral of the (actual) meridional heat flux across a vertical surface of height $H$, stretching the complete latitude circle, which integral was expressed above as the product of the surface and the vertically averaged flux $q=-k^*T_{\phi}/R$. The vertically averaged thermal diffusivity $k^*$ is measured in [W K$^{-1}$ m$^{-1}$]. Its relationship with the diffusion coefficient $k$ can be given as: $k=Hk^*/R^2$.}} as: $j = -\cos(\phi)kT_{\phi}$.}

\section{Methodology}\label{sec:methodology}

\subsection{Model reduction}\label{sec:model_reduction}

The 1d EBM resolves the temperature and so meridional heat transport as functions of latitude, which makes it an infinite degree-of-freedom (DOF) dynamical system. Since we are interested in nonequilibrium thermodynamics in association with heat transport, we would like to define a simple global measure of it for the purpose of analysis. We achieve this essentially by coarsegraining, in order to arrive at a finite DOF model. {This is a top-down approach to diagnosing complex models, as proposed by \cite{LFR:2011}, or to creating a climate model hierarchy, described in detail by \cite{npg-8-211-2001} and \cite{Saltzman}.}

An extreme form of coarsegraining of this system would eliminate the spatial dependence by global averaging. This would lead to the much studied 0d EBMs. A less extreme form of coarsegraining could be achieved by averaging separately over high and low latitudes. This is a common practice in order to establish a minimal model that is capable of resolving global transport properties. A paradigmatic example of this, {as a product of a bottom-up approach}, is Stommel's box model of the thermohaline circulation of the oceans {\citep{Stommel:1961,Dijkstra:2005}}. This conceptual model involves variables representing the average temperature and salinity of the ocean at low and high latitudes, modeled as reservoirs or `boxes', between which heat and material (salt) is exchanged in proportion with their difference in temperature and salinity. The intensity of transports are also controlled by the ocean dynamics in reality, which is crudely represented by parametrizations involving conductivity coefficients. This picture is straightforwardly transferable to the atmospheric dynamics of our interest. Here we consider a 2-box or 2 DOF model (due to the symmetry to the Equator) given by the following two equations:     

\begin{subequations}\label{eq:2boxmodel}
  \begin{equation}\label{eq:2boxmodel_L}
  \begin{split}
    \hat{c}_L\dot{\hat{T}}_L &= f_L(\hat{T}_L,\hat{T}_H)= -\hat{k}(\hat{T}_L-\hat{T}_H) \\
    &+ \mu Q_L[1 - \hat{\alpha}_L(\hat{T}_L)] \\
    &- \hat{\sigma}_L \hat{T}_L^4[1 - m\tanh(c_3 \hat{T}_L^6)], 
  \end{split}
  \end{equation}
  \begin{equation}\label{eq:2boxmodel_H}
  \begin{split}
    \hat{c}_H\dot{\hat{T}}_H &= f_H(\hat{T}_L,\hat{T}_H)= \hat{k}(\hat{T}_L-\hat{T}_H) \\
    &+ \mu Q_H[1 - \hat{\alpha}_H(\hat{T}_H)] \\
    &- \hat{\sigma}_H \hat{T}_H^4[1 - m\tanh(c_3 \hat{T}_H^6)],
  \end{split}
  \end{equation}
\end{subequations} 
where $\hat{T}_L$ and $\hat{T}_H$ represent the average temperatures at low and high latitudes, respectively. On the right-hand-sides, we choose the same functional forms for the radiative terms as in the 1d EBM, so as to express the same physics. (This is indeed the practice also when deriving a 0d EBM.) The terms representing heat transport, identical in the two equations (\ref{eq:2boxmodel_L}) and (\ref{eq:2boxmodel_H}) except for a change of sign, are not obtained at this level of the coarsegraining by a numerical discretization scheme for differentiation, but rather it is modeled as the heat transport between two heat baths between which the temperature varies linearly along the length of the conducting 1-dimensional medium. The transport terms realize a coupling between the two equations. The parameters of the model do not depend on the latitude; we will specify them after a more general discussion of the model.  

In the 2-box model the bistability is preserved, and therefore there exist two branches of a heteroclinic trajectory in the phase plane, linking the unstable saddle fixed point with the two stable node fixed points. A relatively strong coupling causes trajectories which are initialized well between $[T_W]$ and $[T_C]$ to be quickly attracted to the heteroclinic trajectory, `landing' on it at a point not far from the initial $[T_0]$, for which reason the heteroclinic trajectory constitutes a 1-DOF backbone of the transient dynamics. The evolution along this unique trajectory can be described by a 1-DOF model, in terms of either $\hat{T}_L$ or $\hat{T}_H$, or derived variables, such as e.g. the average $\hat{T}=(\hat{T}_L+\hat{T}_H)/2$ or the difference $\Delta \hat{T} = \hat{T}_L-\hat{T}_H$. The latter two together are defined by a nondegenerate transformation of variables. The respective model equations read as follows:

\begin{subequations}\label{eq:1dof_model_LH}
  \begin{equation}\label{eq:1dof_model_L}
    \dot{\hat{T}}_L = f_L^*(\hat{T}_L) = f_L(\hat{T}_L,\hat{T}_H(\hat{T}_L)),
  \end{equation}
  \begin{equation}\label{eq:1dof_model_H}
    \dot{\hat{T}}_H = f_H^*(\hat{T}_H) = f_H(\hat{T}_L(\hat{T}_H),\hat{T}_H),
  \end{equation}
\end{subequations} 
or

\begin{subequations}\label{eq:0dof_models_T_DT}
  \begin{equation}\label{eq:0dof_model_T}
    \dot{\hat{T}} = f_{\hat{T}}^*(\hat{T})=f_{\hat{T}}(\hat{T},\Delta \hat{T}(\hat{T})), 
  \end{equation}
  \begin{equation}\label{eq:0dof_model_DT}
    \dot{\Delta\hat{T}} = f_{\Delta\hat{T}}^*(\Delta\hat{T})=f_{\Delta\hat{T}}(\hat{T}(\Delta \hat{T}),\Delta \hat{T}).
  \end{equation}
\end{subequations} 
Equation (\ref{eq:0dof_model_T}) is nothing but a `standard' 0d EBM when $f_{\hat{T}}^*(\hat{T})$ involves the same functional forms of the radiative terms as e.g. Eq. (\ref{eq:2boxmodel_L}):

\begin{equation}\label{eq:0dEBM}
  \begin{split}
  f_{\hat{T}}^*(\hat{T}) &= \mu \hat{Q}[1 - \hat{\alpha}(\hat{T})] \\
  &- \hat{\sigma} \hat{T}^4[1 - m\tanh(c_3 \hat{T}^6)].
  \end{split}
\end{equation} 
Equation (\ref{eq:0dof_model_DT}) can also be considered a 0d EBM, but it describes a different aspect of the climate: the thermodynamic disequilibrium. However, this model is well-defined only if the function $\Delta \hat{T}(\hat{T})$ is invertible. In fact it turns out to be not the case concerning the energy budget of the climate, and so $f_{\Delta\hat{T}}^*(\Delta\hat{T})$ does not exist, or, it is not a single-valued function. Therefore, it is only the 2-DOF model that can involve $\Delta \hat{T}$ as a prognostic variable.

The constraint that the heteroclinic trajectory imposes is used in both Eqs. of (\ref{eq:1dof_model_LH}) and (\ref{eq:0dof_models_T_DT}), which can be seen as a constitutive relationship between e.g. $\hat{T}$ and $\Delta \hat{T}$, formally written as $\hat{h}(\hat{T},\hat{\Delta T})=0$, to which the 2-DOF dynamics quickly adjusts to a good approximation. Accordingly, $\hat{h}(\hat{T},\hat{\Delta T})\approx f_{\Delta\hat{T}}(\hat{T},\Delta \hat{T})$. The same applies to the 1d EBM: The strong diffusivity quickly brings the trajectory in the infinite-dimensional phase space to the `1-DOF' heteroclinic trajectory, {but so that} shorter length scale ripples in the temperature profile $T(x,t)$ are smoothed out quicker than ripples of longer scales\footnote{{\cite{Ghil:1976} showed that the EBM linearized around a steady state solution can be written in the form of a Sturm-Liouville equation, and that, despite the singularity at the pole, results of the Sturm-Liouville theorem for self-adjoint operators applies, namely that there is a monotonic sequence of {\em real} eigenvalues increasing to (negative) infinity, and in association with these ordered eigenvalues the eigenfunctions have {\em monotonically} increasing number of roots. The diffusive operator alone, the original one less the radiative terms, namely, the Laplace-Beltrami operator on the sphere, has in fact {\em only negative} eigenvalues. These properties dictate the monotonically increasing decay rate of shorter scale ripples. Together with the radiative terms there is therefore at most one positive eigenvalue, and so the largest eigenvalue solely determines the stability, as Ghil points out. We note that these properties are signs of the simplicity of the diffusive model of meridional transport; a more complex realistic `dynamical' transport process would perhaps imply more than one positive Lyapunov exponents, among other new features.}}. Ultimately the global temperature contrast adjusts the slowest. It can be defined as follows:

\begin{equation}\label{eq:DT}
  \Delta T=T_L-T_H, T_L=[T]_0^{1/3}, T_H=[T]_{1/3}^1,
\end{equation} 
where we used the formalism for the area-weighted integral:

\begin{equation}\label{eq:average}
  [T(x,t)]^{x_H}_{x_L} = \frac{\pi}{2}\int_{x_L}^{x_H}dx\cos(\pi x/2)T(x,t).
\end{equation} 
For the definition of $\Delta T$ in (\ref{eq:DT}) we defined the $L$ and $H$ boxes (with a `border' at 30$^\circ$N, i.e., $x=1/3$) so as to represent equal areas. With this partition the resulting boxes experience a heat flux between them which is reasonably close to the maximal value. The actual partition with which the maximal heat flux is achieved is that with a boundary where the net radiative heating rate is zero. As for the equilibria in the full range of bistability with $\alpha_{max}=0.6$ we find this point to be between $x=0.3$ and 0.4, which is in good agreement with an observation of the zonal mean maximum meridional temperature gradient for annual mean conditions occurring at about $37^{\circ}$N \citep{Stone:1978}. We note that  $[T(x,t)]_0^1$ is thus the global average, which we will denote more briefly in the following as $[T]$. The average $[T]$ and the contrast or difference $\Delta T$ can be the first two variables of an infinite sequence that spans the phase space of the 1d EBM. Therefore, the 1-DOF heteroclinic orbit implies a relationship between them:

\begin{equation}\label{eq:constitutive_h}
  h([T],\Delta T;\mu)=0.
\end{equation} 

Now we pick up with the discussion of the parametrization of the finite-DOF models. 
The right-hand-side $f_{\hat{T}}^*(\hat{T})$ of Eq. (\ref{eq:0dEBM}) implicitly assumes a parametrization which realizes the very same solution for the 1-DOF model as the heteroclinic orbit of the 2-DOF model. Furthermore, we also require that the irradiation and heat loss terms, respectively, in Eqs. (\ref{eq:2boxmodel}) sum up separately to those of the corresponding terms of Eq. (\ref{eq:0dEBM}). It is only possible if $\hat{\sigma}=\hat{\sigma}(\hat{T})$. This idea can be transferred as for the parametrization of the 2-DOF model in order to have $h(\cdot,\cdot;\cdot)=\hat{h}(\cdot,\cdot;\cdot)$, and also that the individual terms (tendency of internal energy, insolation, heat loss to space) of the $L$ and $H$ components of the 2-DOF model equal exactly the $[\cdot]_0^{1/3}$ and $[\cdot]_{1/3}^1$ integrals, respectively, of the corresponding terms of the 1d EBM. This will require the parameters to depend on the system state -- and also on system parameters of interest, e.g.: $\hat{c}_H=\hat{c}_H(\hat{T};\mu)$, $\hat{k}=\hat{k}(\hat{T};\mu)$, $\hat{\alpha}_H=\hat{\alpha}_H(\hat{T};\mu)$, $\hat{\sigma}_H=\hat{\sigma}_H(\hat{T};\mu)$ (and $Q_H=[Q(x)]_{1/3}^1$). In order to determine the functions $\hat{c}_H(\hat{T};\mu)$, etc., the 1d EBM has to be solved. As an implication, the two eigenvalues of the 2-box model, concerning any of the equilibria, are identical to the first two eigenvalues of the 1d EBM. In particular, the positive eigenvalue of the unstable equilibrium of the 1d EBM equals the corresponding eigenvalue of the 0d EBM (\ref{eq:0dof_model_T}). However, in a strongly diffusive model the diffusive term has a relatively small contribution to this eigenvalue. {Appendix \ref{apdx:phase_portrait} sheds more light onto the relationship of the finite- and infinite-DOF models, and also on the concepts of dynamical systems theory that we used here, such as those of the phase space, basin boundary, heteroclininc orbit, eigenvector, etc.}

For our purpose, to determine the constitutive relationship (\ref{eq:constitutive_h}), we do not need to actually determine such a parametrization of the 2-box model, but we can proceed directly by integrating Eq. (\ref{eq:pde}). The 2-box model is intended here as a device to interpret our results, and to draw a connection with the literature.

Another common measure of thermodynamic imbalance is the material entropy production in the process of down-gradient heat transport. In case of the 2-box model, assuming that the rate of irreversible heat transfer between two heat baths $J=\hat{k}\hat{\Delta T}$, the following formula known \citep{GM:1969} for thermal resistors applies:

\begin{equation}\label{eq:2box_smat}
  \dot{\hat{S}}_{mat} = J\left(\frac{1}{T_H}-\frac{1}{T_L}\right) \approx \hat{k}(\hat{T})(\hat{\Delta T}/\hat{T})^2.
\end{equation}  
This approximation of the entropy production is a diagnostic quantity for the 2-box model, and it provides an alternative to $\Delta \hat{T}$ for the definition of a climatic constitutive relationship similar to (\ref{eq:constitutive_h}). Concerning the dynamics restricted to the heteroclinic orbit, $\dot{\hat{S}}_{mat}\approx\dot{S}_{mat}$ to a very good approximation, the material entropy production in the 1d EBM being defined~\citep{Paltridge:1978,Grassl:1981} as:

\begin{equation}\label{eq:dsdt_mat}
    \dot{S}_{mat} = A\int_{0}^{\pi/2}d\phi \cos(\phi)k(\phi,T)(T'_{\phi}/T)^2,
\end{equation} 
where $A$ is the surface of the globe.

\subsection{Finding unstable solutions: Edge tracking}\label{sec:edgetracking}

In order to determine the constitutive relationship (\ref{eq:constitutive_h}), we need to find the unstable states. As mentioned before, one possible approach would be through solving the relevant boundary value problem. It is yet to be seen whether it is a feasible task. Another approach is to consider the unstable climate as the solution of an initial value problem. When the objective is to find an attracting steady state, an attractor, choosing initial conditions is usually straightforward, as any initial condition, even that of uniform pressure, temperature, etc. fields, from within its basin of attraction is attracted by the steady state. In contrast, an unstable state does not have a basin of attraction, only a measure zero set of initial conditions would converge to it, which initial conditions in fact consist of the stable manifold of the unstable state; even close-by initial conditions to the stable manifold will be repelled upon unconstrained forward-integration of the governing differential equations. Nevertheless, when e.g. the unstable state is due to bistability, a straight line in a possibly high-dimensional phase space -- that connects two initial points that lead to different attractors -- intersects the stable manifold (a surface) with probability one. An efficient numerical technique, called edge tracking, which involves the bracketing of the said intersection point with the stable manifold will be our choice for finding unstable states. It was first proposed and applied for finding unstable solutions of shear flow problems -- referred to as edge states -- by~\cite{PhysRevLett.96.174101}, and is described in more detail by \cite{Tobias:2011}. The main objective of this paper is to describe this technique in detail and demonstrate its applicability to a geophysical system, namely, the {Ghil-Sellers} 1d EBM described in Sec. \ref{sec:model}. We wish also to summarize practical considerations regarding the algorithm, such as its robustness, efficiency, and accuracy, with a view to applying the technique to more complex climate models.  

\begin{figure}
\centerline{\includegraphics[width=\linewidth]{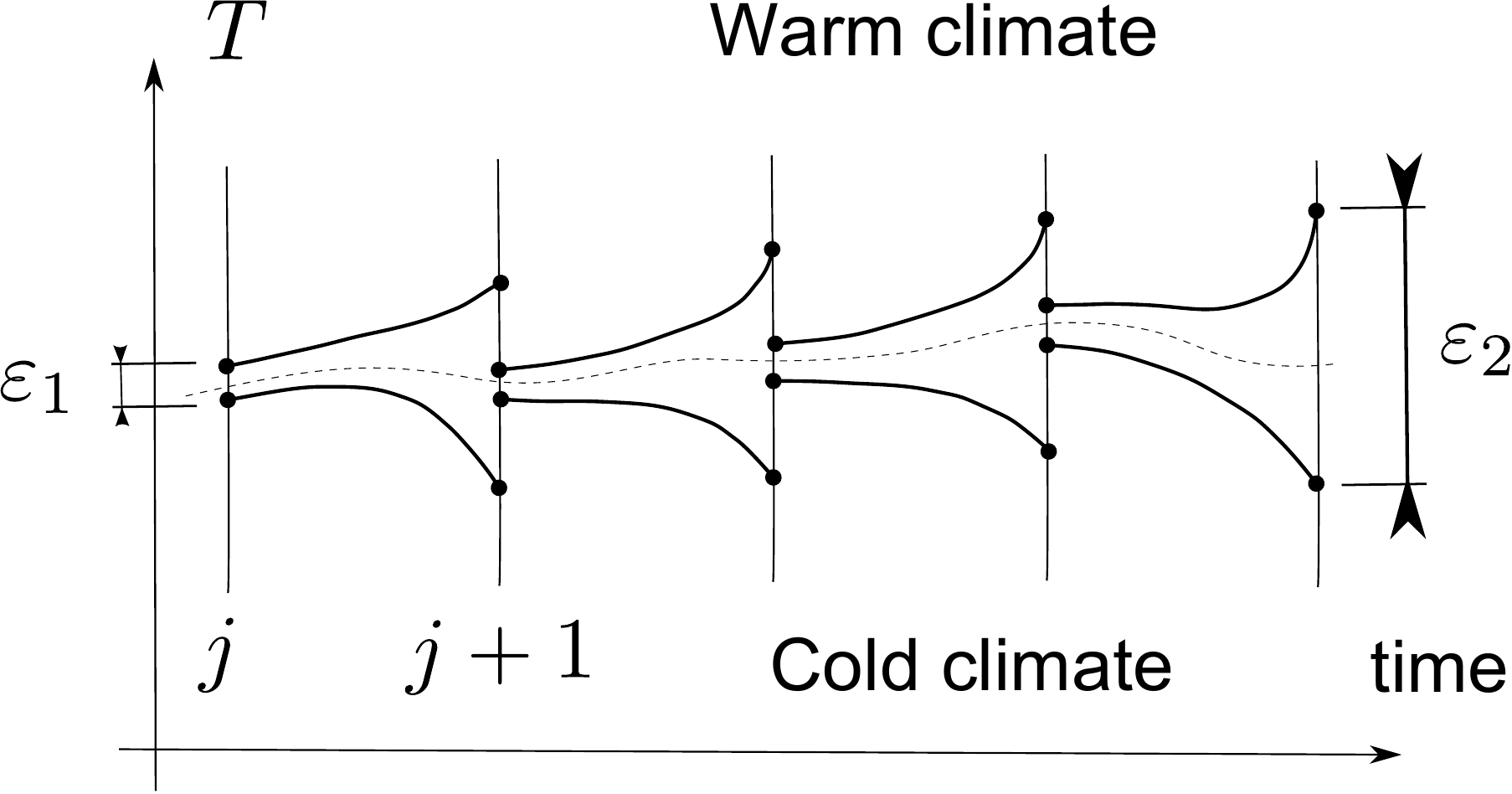}}
\caption{Schematic illustration of the edge tracking algorithm. The dashed line represents the evolution of the unstable state $T_U(t)$ in terms of a scalar bulk quantity (the symbol $T$ refers to the intuitive spatially averaged temperature), and pairs of bracketing trajectories drawn by solid lines are shown to diverge from it. Vertical straight lines mark the times of reinitialization, explained in the main text. Note that the divergence of the trajectories is blown up for visualization to exceed the variance of the unstable state, $\varepsilon_2 > \textrm{var}[T_U(t)]$, whereas the opposite, $\varepsilon_2 < \textrm{var}[T_U(t)]$, is necessary for a good signal-to-noise ratio for resolving $T_U(t)$.} \label{fig:et_illustration}
\end{figure}

Before a detailed description of the edge tracking technique, we give a brief illustration of it based on the schematic in Fig. \ref{fig:et_illustration}, where the evolution of the unstable state is represented by the dashed line. To track an unstable state by forward-integration, the integration has to be repeatedly stopped and reinitialized, because the numerical trajectory is repelled from the unstable state. This situation is shown in Fig. \ref{fig:et_illustration}, where pairs of bracketing trajectories, drawn out by solid lines, diverge from the dashed line. In case of a 0d EBM we have a simple picture of this behavior: the unstable equilibrium is situated on top of a potential barrier, at the local maximum of the double-well potential function (which may be forced to vary over time), and the solid lines would evolve towards the bottom of one of the potential wells. 

Still with a reference to the schematic in Fig. \ref{fig:et_illustration}, when two trajectories are initialized on the two sides of the unstable equilibrium with a very small separation ($\varepsilon_1$), they stay close -- within a finite window containing the unstable equilibrium -- for a long time, with a lifetime that `blows up' with vanishing separation. In fact the length of the lifetime can be the basis for bracketing closely the unstable equilibrium again once the two trajectories have diverged to a distance larger than a threshold ($\varepsilon_2$), which is the core idea of the PIM-triple algorithm \citep{Nusse1989137}. However, when the unstable solution is due to bistability, its bracketing can be done more efficiently by the edge tracking algorithm performing an iterative bisecting procedure by checking which stable state an intermediate trajectory evolves towards, as described in detail next. The algorithm is applicable to higher-DOF systems too, in which case what is bracketed is the basin boundary. Embedded in it is the unstable state, which may be a simple object like a stable node as in case of the diffusive 1d EBM, or a set of a more complex geometry. In forthcoming studies we wish to apply this algorithm to discover the unstable climate in the GCM PlaSim.

For our analysis we implemented the algorithm in Matlab by using \texttt{pdepe}. Each cycle ($j$) of the algorithm has the two main phases of advancing and reinitializing the system:  

I. {\em Iterative bisecting}. In the initial cycle ($j=0$) we start out with two temperature profiles, $T_{w,i,j}(x)$ and $T_{c,i,j}(x)$, $i,j=0$, which lead to the two different coexisting stationary climates, and consequently this property will be given later for any of $j=1,\cdots,J$ following from the nature of the algorithm. With the iterative bisecting ($i=0,1,2,\dots$ for any fixed $j$) an intermediate profile is searched for, from which it takes a relatively long time to end up with one of the stationary climates. In this phase in fact three actions are iterated. First ($j=0$) (a) we linearly interpolate between the warm and cold climates, conveniently -- and later on ($j>0$) between the warm- and cold-side profiles:

$$T_{m,i,j}(x) = [T_{w,i,j}(x) + T_{c,i,j}(x)]/2.$$
Then (b) we check by forward simulation if $T_{m,i,j}(x)$ leads to the warm or cold climate. If it is the former case, then (c) the procedure is reinitialized 
such that $T_{w,i+1,j} = T_{m,i,j}$, and otherwise $T_{c,i+1,j} = T_{m,i,j}$. 

In order to check which scenario is realized, we monitor a suitable scalar indicator quantity in the course of the forward simulation {in phase I.b}, and check if an upper or a lower threshold is exceeded. These thresholds are determined as a preliminary exercise, verifying that their excedance predicts the outcome with certainty. Here, with the 0d EBM and the schematic illustration in Fig. \ref{fig:et_illustration} in mind, the area-weighted mean temperature $[T(x)]_0^1$ is perhaps the most intuitive indicator. The thresholds can be chosen to be in the close vicinity of the asymptotic value, because the approach of any of the stable solutions is relatively fast, and so the outcome cannot be predicted with significantly less simulation effort. This implies that -- from a merely technical point of view -- any quantity will be suitable for the indicator purpose, if it takes appreciably distinct values near the two attractors. From a theoretical point of view,  the indicator quantity is desired also to give an insight to the physical mechanism at work. For finding edge states in pipe flow the turbulent energy was taken as an indicator quantity by \cite{PhysRevLett.96.174101}. The turbulent energy in the edge state is intermediate between the transient turbulent and the stable laminar states. 

Further possible indicator quantities in the context of the 1d EBM are the following. The latitude of some high albedo, e.g. $x|_{\alpha=0.5}$ indicates the extent of the snow cover, and therefore it is closely related to the average temperature. As we wish to study both the structural properties and transient dynamics from the point of view of thermodynamic disequilibrium, various measures of it, e.g. $\Delta T$ and $\dot{s}_{mat}=\dot{S}_{mat}/A$, are also sensible choices for indicators.

Actions I.(a-c) are repeated ($i=0,1,2,\dots,I$) while, say, $[T_{w,i+1,j} - T_{c,i+1,j}]_0^1 > \varepsilon_1$, that is, while the reinitialized warm and cold-side profiles differ -- in terms of e.g. the mean $[\cdot]_0^1$ -- from each other by more than a prescribed small value $\varepsilon_1$. In effect by $T_{m,I,j}$ ($T_{c,I,j}$ and $T_{w,I,j}$) we have closely approximated (bracketed) the stable manifold of the unstable solution in this phase. 

II. {\em Advancing}. The next step is to let $T_{m,I,j}$ evolve under the dynamics towards the unstable solution along its stable manifold: $\mathcal{E}_t[T_{m,I,j}]$, where $\mathcal{E}_t[\cdot]$ denotes the nonlinear (autonomous) evolution operator advancing the initial condition up to time $t$. However, by a straightforward simulation the evolution has also an unstable component besides the stable one. In fact $T_{w,I,j}$ and $T_{c,I,j}$ are also advanced, which results in the separation of their trajectories. For this reason the advancing of the profiles is stopped at time $t_{j+1}$ when $[\mathcal{E}_{t_{j+1}}[T_{w,I,j}] - \mathcal{E}_{t_{j+1}}[T_{c,I,j}]]_0^1 = \varepsilon_2 > \varepsilon_1$, using some small $\varepsilon_2$. 

{\em Robustness.} In order to ensure its {robustness}, a third (III) phase of the procedure adjusting the pair of bracketing trajectories might be necessary \citep{Tobias:2011} when one simulates the system using -- otherwise efficient -- adaptive time step and/or implicit numerical integrator schemes. This phase is described in Appendix \ref{apdx:phase_III}.

{\em Convergence.} In the case of the autonomous 1d EBM (\ref{eq:pde}) when the unstable climate is time-independent, the procedures in phases I.-III. can be repeated a few times ($j=0,1,2,\dots,J$) in order to refine the approximation of the unstable solution, which is actually achieved in phase II. The rate of {convergence} to the unstable state is governed by its largest negative eigenvalue. (That is, strictly speaking, the property of convergence is not a property of the algorithm but that of the treated system.) Since the 1d EBM is strongly diffusive, its largest negative (second) eigenvalue in modulus $|\lambda_U^{(2)}|$ is large, and so the convergence is fast. 

{\em Numerical efficiency.} When the unstable state is time-dependent, including the realistic scenario of a chaotic unstable state, the procedures can be continued indefinitely ($j=0,1,2,\dots$) in order to produce an arbitrarily long {\em numerical} trajectory. In general, at any stage of the procedure, i.e., for any $j$, $\varepsilon_1$ can be chosen arbitrarily small for the iterative bisecting (phase I) in order to approximate the stable manifold, and so the unstable state, arbitrarily close, which would entail the desirable property of an arbitrarily long trajectory lifetime $t_{j+1}-t_j$. To do this in practice is not efficient, however. To see why, we have to assess the bracket size $\varepsilon_1$-dependence of the computational effort that is needed to produce a given length of numerical trajectory. The length of the (cold or warm-side) trajectory ($LT=t_{j+1}-t_j$) in a single $j$-cycle ($j>0$) that approximates the stable manifold in the beginning ($t_j$) with a distance of {\em at most} $\varepsilon_1$ can be found as follows:

\begin{equation}\label{eq:LT}
  LT(\varepsilon_1,\varepsilon_2) = I\ln2/\lambda_U^{(1)},
\end{equation} 
where $\lambda_U^{(1)} = df_{\hat{T}}/d\hat{T}|\hat{T}_U$ is the positive largest eigenvalue of the 1d EBM, and the parameter-dependence is introduced through the necessary number of bisections $I=\lceil(\ln\varepsilon_2 - \ln\varepsilon_1)/\ln2\rceil$. In the latter the ceiling function $\lceil x\rceil$ gives the smallest integer not smaller than $x$. For $j=0$ the larger bracket size is not $\varepsilon_2$, but something else depending on the initialization of the procedure, e.g. $[T_W-T_C]$. Equation (\ref{eq:LT}) is valid in the linear limit using small enough $\varepsilon_2$. Next, the computational effort ($CE$) needed in the same $j$-cycle in units of model time -- so as to be comparable to $LT$ -- can be shown easily to be:

\begin{equation}\label{eq:CE}
    CE(\varepsilon_1,\varepsilon_2) = LT(I+1)/2 + t_{I.b}I.
\end{equation} 
The first term accounts for the $I$ cycles of the iterative bisecting, expressing the length of bracketing trajectories diverged to an $\varepsilon_2$ degree, and the second term for the same additional time $t_{I.b}$ in each iteration needed to cross the prescribed threshold in order to determine (in phase I.b) which attractor a trajectory would go to. Note that, first, phase II is also accounted for; second, we assumed stringent enough error tolerance so that phase III is not needed, as mentioned in Appendix \ref{apdx:phase_III}; and third, we can neglect other computational efforts (use of auxiliary variables, checking conditions, etc.) done per cycle. Therefore, we have that

\begin{equation}\label{eq:CEperLT}
  CE/LT = (I+1)/2 + t_{I.b}\ln2/\lambda_U^{(1)}.
\end{equation} 
The relative computational effort $CE/LT$ can be minimized by minimizing $I$. This dictates a single $I=1$ bisection per $j$-cycle, i.e., {\em noniterative} bisection! Translating this into terms of the parameters: given some requirements on the accuracy that fixes $\varepsilon_2$, we need $\varepsilon_2/2<\varepsilon_1<\varepsilon_2$. (Note that a too small $\varepsilon_2$ would not be sensible given that some errors of approximating the continuous-$x$ temperature profile $T(x)$ in the computational gridpoints are already incurred due to the spatial coarsegraining.) This way we piece together a long numerical trajectory from segments (as the schematic in Fig. \ref{fig:et_illustration} suggests) of {\em minimum} length $\ln2/\lambda_U^{(1)}$. Note that in case of a large-DOF system $\lambda_U^{(1)}$ can be thought of as the largest positive finite-$LT$-time Lyapunov exponent (FTLE) of the chaotic saddle, $\lambda_U^{(1,LT)}$ \citep{BKT:2011}. As this quantity controls $LT$, and generally varies over time, $LT$ will also be different in the subsequent $j$-cycles.

\section{Results}\label{sec:results}

\subsection{Edge tracking}\label{sec:results_edge_tracking}

\begin{figure}
\centerline{\includegraphics[width=\linewidth]{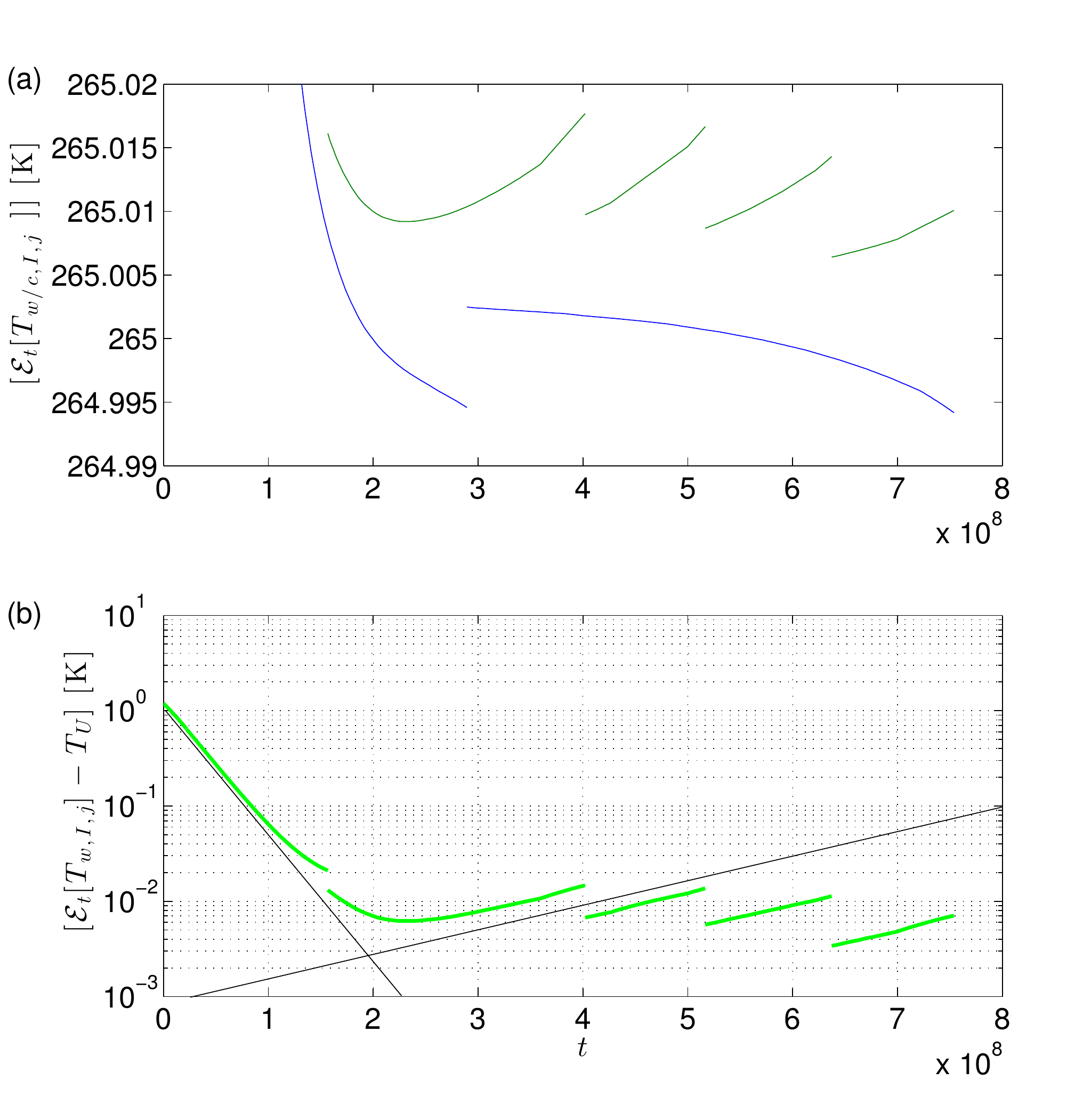}}
\caption{Bracketing trajectories. (a) Pairs of $J=7$ warm and cold-side bracketing trajectories resulting from the edge tracking procedure. (b) Only the warm-side trajectory is shown, having subtracted the constant value of $[T_U]$ estimated to be about 265.003 K, in a lin-log diagram. The rates of exponential convergence and divergence to and from the edge state are given by the slopes of the black straight lines tangential to the trajectory.} \label{fig:bracketing_trajs}
\end{figure}

For $\mu=1$ the bracketing trajectories resulting from the edge tracking procedure are shown by Fig. \ref{fig:bracketing_trajs} (a). We used $[T]$ to determine which stable state is approached. Although we can use any initial conditions from the different basins of attraction, we initialized the warm and cold-side trajectories involved in the edge tracking procedure by the warm and cold stable states: $T_{w,0,0}=T_W$ and $T_{c,0,0}=T_C$. We are yet to choose the tolerance parameters $\varepsilon_1$ and $\varepsilon_2$. This hinges on the fact, partly, that we adopt in our numerics the optimal choice of $I=1$ for $j=1,2,\dots$, which restricts the choice of $\varepsilon_1$ as stated above. In the initial $j=0$ cycle, however, typical initial conditions (not near the stable manifold of the edge state) require more than one bisections: $I_0=\lceil(\ln[T_{w,0,0}-T_{c,0,0}]-\ln\varepsilon_1)/\ln2\rceil$, which realizes a small bracket size $\varepsilon_0=[T_{w,0,0}-T_{c,0,0}]/2^{I_0}$ and also an initial trajectory length $LT_0=(\ln\varepsilon_2-\ln\varepsilon_0)/\lambda_U^{(1)}$. The latter can be different from the constant length of $\ln2/\lambda_U^{(1)}$ in the latter cycles. In particular, $LT_0$ can be arbitrarily small as $\varepsilon_0$ approaches $\varepsilon_2$. We can generate such a situation by choosing $\varepsilon_1$ so that $\varepsilon_1/\{\varepsilon_1-\varepsilon_0)\gg1$, and furthermore choosing $\varepsilon_2$ so that $\varepsilon_2/\{\varepsilon_2-\varepsilon_1)\gg1$, in addition to- and consistent with the above stated restriction. Merely for the purpose of demonstrating the possibility of a short initial trajectory length, accordingly, we chose $\varepsilon_1=1.5\cdot10^{-2}$ [K], and we set $\varepsilon_2 = 1.05\varepsilon_1$ [K], which implies an acceptable accuracy for us. This will not affect our analysis of the results following below, in comparison with a more arbitrary choice of $\varepsilon_1$ according to the restriction $\varepsilon_2/2<\varepsilon_1<\varepsilon_2$. Clearly, the smaller bracket size for $j>0$ is $\varepsilon_2/2$, which is just a little larger than $\varepsilon_1/2$, and so a second bisection is never needed. This is well visible in Fig. \ref{fig:bracketing_trajs} (a), along with the fact that no phase III of the procedure was necessary provided that we required a sufficiently small relative error of integration.

$J=7$ cycles of the procedure have been completed, but we see that the convergence to the edge state is not improved already from the third cycle on. As the rate of convergence to/divergence from the edge state is governed by its largest negative/positive eigenvalue, we can estimate it as a slope of e.g. the warm-side trajectory, once the constant value of $[T_U]$ is subtracted (guessed to be about 265.003 K), and the trajectory is presented in a lin-log diagram, as seen in Fig. \ref{fig:bracketing_trajs} (b). The negative eigenvalue can be approximated by the initial slope, indicated in the diagram by a solid straight black line, provided that $T_{w/c,I,0}$ are relatively close to the edge state and that $\varepsilon_1$ is small enough, which are reasonably satisfied in our case (supported shortly below). The warm-side trajectory is leaning close to the straight line in the short $j=0$-cycle and in part of the $j=1$-cycle. The first discontinuity point of the piece-wise discontinuous numerical trajectory between the first two cycles is not really visible in the lin-log diagram, but the second and so the rest of the discontinuity points are already apparent in our case. We can read off a negative initial slope of about $-3.0\cdot10^{-8}$. The positive slope, say, for $j=5$, also indicated by a straight line, is about $6.1\cdot10^{-9}$. This implies $LT\approx1.15\cdot10^{8}$, in agreement with what can be seen in the diagram. Importantly, the negative slope is about 5$\times$ that of the positive slope, which confirms our previous statement on relatively strong diffusivity and fast convergence, something that underpins our concept of the constitutive relationship (\ref{eq:constitutive_h}). Moreover, since the minimum $LT$ is controlled by the positive slope, it is clear now that the edge state in our case is closely approximated already in the first, or at most the second, cycle ($j=0,1$); and so the slope after that does represent the positive eigenvalue. We have obtained the first two eigenvalues of the (nonlinear) 1d EBM (\ref{eq:pde}) by solving the related characteristic equation, by using \texttt{bvp4c} once more. The method is detailed by \cite{Ghil:1976}. The resulting values, $\lambda_U^{(1)}=6.84\cdot10^{-9}$ and $\lambda_U^{(2)}=-2.34\cdot10^{-8}$, are in reasonable agreement with the estimates given by the slopes. 

We note that the `mismatch' of these eigenvalues is related to the steepness of the eigenvector belonging to $\lambda_U^{(2)}$ projected onto the $[T]$-$\Delta T$ plane. Trajectories approach closely the edge state or the heteroclininc trajectory running approximately in parallel with this eigenvector; and since the latter makes a small angle with the vertical, we do see a fast initial evolution of $[T]$ governed by $\lambda_U^{(2)}$.

Subsequent $j$-cycles of the edge tracking procedure applied to the 1d EBM refine the {\em shape} of the unstable equilibrium temperature profile $T_U(x)$ more and more. Otherwise, the approximation of the {\em mean} by e.g. the warm-side trajectory $[T_{w,I,j}]$ can be shown to be a discrete-time chaotic process. Due to this fact the bisection affects the cold and warm-side trajectories, regarding whether they are continued or reinitialized, with the same relative frequency. Continued application of the edge tracking procedure, upto a much larger $J$ than in Fig. \ref{fig:bracketing_trajs} (a), would show this. Properties of the bisection, however, as discussed above, guarantee that $[T_{w,I,j}-T_U]<\varepsilon_1$ for all $j$ and any choice of $\varepsilon_2$.

In the general case of a time-dependent edge state, however, the continuous-time evolution of e.g. $|[\mathcal{E}_t[T_{w,I,j}]-T_U(t)]|$ for any fixed $j$ is not expected to be monotonically increasing, and $|[T_{w,I,j}-T_U(t_j)]|$ might be of order $\varepsilon_2$. Although, this would not jeopardize the control of {\em accuracy} by $\varepsilon_2$.

Considering more complex GCMs featuring bistability, we expect that the outlined edge tracking procedure is still applicable. A bisection would simply constitute a linear interpolation between two points that belong to different basins of attraction in a large $d$-dimensional phase space. The straight line (of dimension one) that connects the two points will then have one or more intersection points (of dimension zero or less than one) with the possibly folded stable manifold of the unstable state, whose dimension is at least $d-1$. From such an intersection point, in the general case of a time-dependent edge state, the convergence of the numerical trajectory is governed by the largest negative `local' or time-dependent FTLE, and the lifetime of trajectories in the $j$-cycles are controlled by the largest positive time-dependent FTLE. The coexisting cold and warm states of PlaSim, for example, are chaotic states. Whether the corresponding unstable state is a fixed point, a periodic orbit, a chaotic saddle, or some other object, is yet to be seen.

\subsection{Transient dynamics and functional relationships between observables}

The backbone of the transient dynamics is the heteroclinic trajectory in the 1d diffusive EBM. This means that an {\em arbitrarily} initialized trajectory (but again: initialized well between $[T_W]$ and $[T_C]$) would be first quickly attracted to the heteroclinic trajectory, and then it would slowly `slide' to the stable state confined closely to that trajectory. The direction of the quick initial approach of the heteroclinic trajectory is given by the eigenvector belonging to $\lambda_U^{(2)}$, which is the direction of the slowest approach \citep{TG:2006}, but it is still quick in a strongly diffusive system ($\lambda_U^{(1)}\ll|\lambda_U^{(2)}|$).

\begin{figure*}
\centerline{\includegraphics[width=\linewidth]{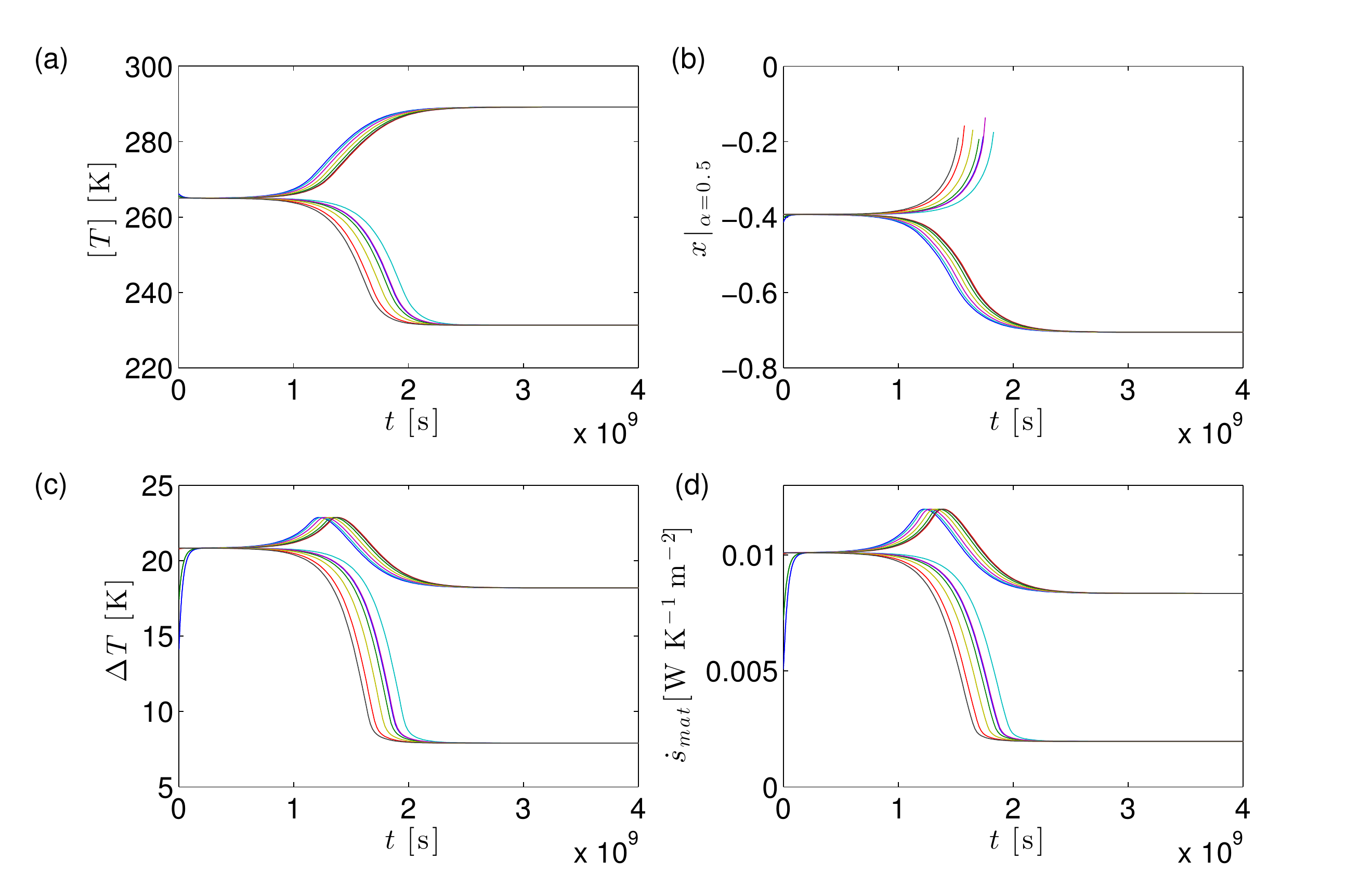}}
\caption{The evolution of various possible indicator quantities in phase II of the edge tracking, extended all the way to the stable states: (a) average temperature $[T(x,t)]_0^1$ (b) latitude of an intermediate value of the albedo $x|_{\alpha=0.5}$ (c) bulk temperature difference between high and low latitudes $\Delta T$ (d) (specific) material entropy production in the process of meridional heat transport $\dot{s}_{mat}$. $J=7$ (color matching) pairs of warm- and cold-side time series for each quantity are shown (although note that the orders of the warm- and cold-side time series do not have to be the same for any pair of the indicator quantities).} \label{fig:indicator_quantities}
\end{figure*}

The heteroclinic trajectory can be obtained as a `byproduct' of the edge tracking procedure in a way that the advancing in phase II is continued even after the $\varepsilon_2$ degree of separation of the bracketing trajectories. That is, the trajectories seen in Fig. \ref{fig:bracketing_trajs} (a) can be extended all the way to the stable states. Figure \ref{fig:indicator_quantities} shows the extended $J=7$ pairs of trajectories in terms of the time evolution of the four possible indicator quantities named in Sec. \ref{sec:methodology}. Note that for this picture time is reinitialized to 0 in every new cycles of the edge tracking procedure. We emphasize that it does not matter which of these quantities is actually used for the indicator purpose in the procedure, they all yield the same unstable state. It is shown by hollow circle markers in Fig. \ref{fig:temp_profiles} (a) alongside the solution found by Matlab's boundary value problem solver \texttt{bvp4c}, indicating a very accurate result. Equilibrium values of the four indicator quantities are compiled in Table \ref{tab:indicator_values}.

\begin{table}
\caption{Values that four indicator quantities (described in the main text) take for the warm (W), cold (C), and the unstable (U) climates. 
}
\begin{center}
\begin{tabular}{l|llll}
 & $[T]$ [K] & $x|_{\alpha=0.5}$ & $\Delta T$ [K] & $\dot{s}_{mat}$ $\left[\frac{\textrm{mW}}{\textrm{K} \textrm{m}^{-2}}\right]$\\
\hline
W & 289.0 & 0.70 & 18.2 & 8.5\\
U & 265.0 & 0.39 & 20.8 & 10.1\\
C & 231.3 & $(\alpha=0.6)$ & 7.9 & 2
\end{tabular}
\end{center}
\label{tab:indicator_values}
\end{table}

\begin{figure*}
\centerline{\includegraphics[width=\linewidth]{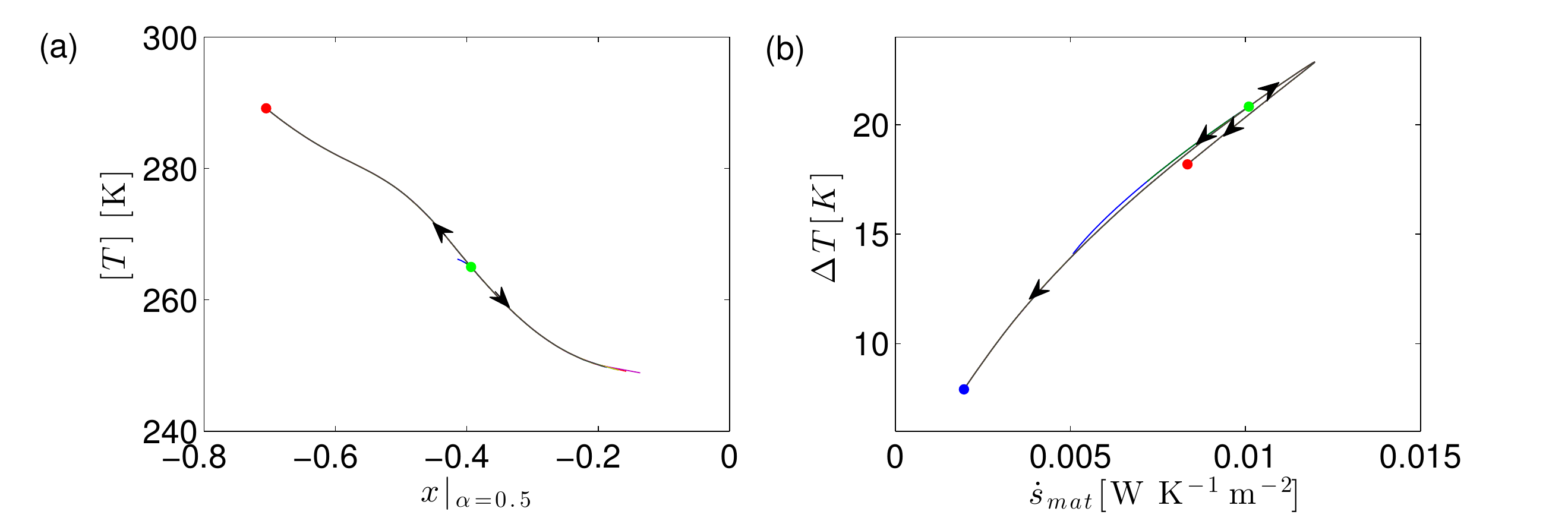}}
\caption{Relationships between observables used as indicator quantities for the edge tracing procedure. Pairs of time series of (a) $[T]$ vs $x|_{\alpha=0.5}$ or (b) $\Delta T$ vs $\dot{s}_{mat}$ from Fig. \ref{fig:indicator_quantities} are plotted together. The position of equilibria are indicated by circular markers. (Note that the cold state could not be defined in terms of $x|_{\alpha=0.5}$, because $\alpha=0.6$ everywhere on the snowball Earth.) Arrow heads indicate the direction of change along branches of the heteroclininc orbit, pointing away from the unstable state.} \label{fig:comparisons}
\end{figure*}

When any two indicator quantities are plotted against one another, the $J=7$ curves collapse onto one -- as they should, all being closely confined to the same unique heterclinic trajectory. This establishes a functional relationship between the pair. We paired  up $[T]$ vs $x|_{\alpha=0.5}$ and $\Delta T$ vs $\dot{s}_{mat}$ in Fig. \ref{fig:comparisons} (a) and (b). As for the first pair, since $x|_{\alpha=0.5}$ is a reasonable proxy for the (long-time average) ice cap extent, its close relation to $[T]$ is expected indeed. Previous studies also considered the ice cap extent to determine the tipping point quantitatively. Our figure of $x|_{\alpha=0.5}=0.39$ for the unstable state is in reasonable agreement with the finding of \cite{DRFRG:2004} in a coupled ocean-atmosphere model of intermediate complexity, such that when the rather sharp crossover of the mean annual see ice cover fraction from 1 to 0 reaches the latitude of about 30$^{\circ}$ North, the model climate switches abruptly to a snowball state. Therefore, if $[T]$ is governed approximately by a well-defined 0d EBM, then so is $x|_{\alpha=0.5}$ (at least in a limited range, which does not include the stable cold completely snow covered snowball state). Consequently, if the unstable state is slightly perturbed toward the cold state, then the climate experiences a monotonic or gradual increase of the snow cover; and similarly, a perturbation towards the warm state would initiate a gradual receding of the snow cover.

Considering the other pair of observables, the functional relationship between $\Delta T$ and $\dot{s}_{mat}$ is also rather obvious, warranted by the approximation (\ref{eq:2box_smat}). However, in Fig. \ref{fig:comparisons} (b) we observe a sharp turn of the curve on the warm-side branch, which is a not so obvious effect. It is related to the fact that e.g. $\Delta T$ cannot be simulated by a surrogate 0d EBM, i.e., a single-box model, as mentioned in Sec. \ref{sec:model_reduction}. This is reflected in the transient increase and nonmonotonic evolution of $\Delta T$ and $\dot{s}_{mat}$ towards the stable warm state shown in Fig. \ref{fig:indicator_quantities} (c) and (d), respectively. That is, the climate experiences the state of most active heat transport at an intermediate out of dynamical equilibrium state, between the unstable and warm stable equilibria. 

\begin{figure*}
\centerline{\includegraphics[width=\linewidth]{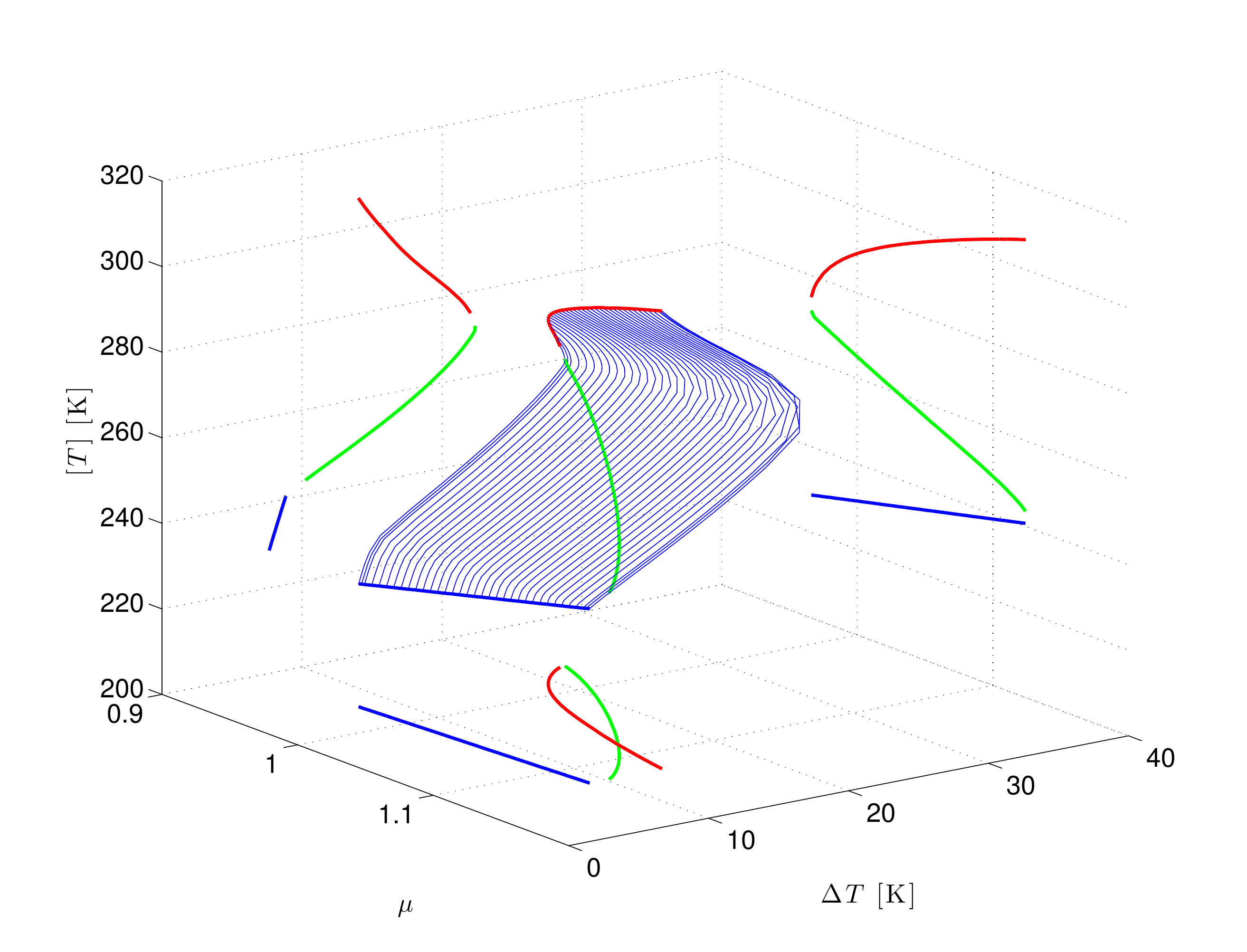}}
\caption{Constitutive relationship between $\Delta T$ and $[T]$ (depending only slightly on the solar strength $\mu$) corresponding to the heteroclinic trajectory to which the system very quickly aligns. The surface $h([T],\Delta T,\mu)=0$ is mapped out by thin solid (blue) lines; and  in the surface thick solid lines mark the steady states. Projections of these latter curves onto planes spanned by the different combinations of two of the three variables are also shown.} \label{fig:constitutive_rel}
\end{figure*}

It is useful to construct the functional relationship also between $\Delta T$ and $[T]$, which we called a constitutive relationship in Sec. \ref{sec:model_reduction}. Figure \ref{fig:constitutive_rel} shows this relationship between $\Delta T$ and $[T]$ mapped out for the full range of bistability in terms of $\mu$, which can be represented as a surface: $h([T],\Delta T,\mu)=0$ [refer to Eq. (\ref{eq:constitutive_h})]. A characteristic feature of this surface is that for any fixed $\mu$ the single maximum of $\Delta T$ takes place at about $[T]\approx270$ K in the full range of bistability. Also more generally, the dependence on $\mu$ is not strong. The border of the surface on one side is drawn out by the stable warm states depending on $\mu$ (thick solid red line), and on the opposite side by the stable cold states (thick solid blue line). The 3d parametric curve of any of the equilibria given by e.g. $\{\Delta T(\mu),[T(\mu)],\mu\}$ we will refer to as the path of the equilibrium state in question. The path of the unstable state is shown by the thick solid green line. This path does not align with the ridge of the surface, max$[\Delta T]|_{\mu=const}$. We conclude, therefore, that a transient increase of $\Delta T$ towards the warm state is possible.

In surface $h$ branches of the heteroclinic trajectories belonging to fixed values of $\mu$ can be seen as thin blue lines. Let us compare the transient dynamics for different fixed values of $\mu$. Starting from the unstable state, triggered by a small perturbation towards the warm state, for larger values of $\mu$, which imply smaller $\Delta T$ initially, the system has to `climb' more in order to reach the same maximal $\Delta T$ (which depends much more weakly on $\mu$ in comparison e.g. with the paths of equilibria), and then it would descend to a lower value of $\Delta T$ of the warm state. That is, the larger $\mu$ is, the `lengthier' excursion the system has to take in terms of $\Delta T$. If the system is initialized arbitrarily, it first quickly goes to the (blue) heteroclinic trajectory, and from then it might or might not have to take an excursion in terms of $\Delta T$ in order to arrive at the stable state.

In the general case of time-dependent chaotic unstable states, there is a pair of dense sets of trajectories that lead from the unstable to the two different stable states. A skeleton of the chaotic set is constituted by a dense set of periodic points of the saddle type, each of which has one or more unstable directions. Several types of trajectories may be associated with these unstable directions, and here we are concerned with two types that belong to the unstable manifold of the unstable set. Among these a heteroclininc trajectory `connects' an individual (unstable) periodic point of the unstable set with an/the (unstable) periodic point of one of the attractors. If the attractor is chaotic, another type of trajectory originating from an unstable periodic point (or its immediate vicinity) of the unstable set would approach one of the chaotic attractors and keep winding around it. In summary, a dense set of unstable periodic points of the unstable set are associated with a dense set of unstable directions and associated trajectories. Therefore, a constitutive relationship similar to (\ref{eq:constitutive_h}) can be defined only in a statistical sense, as an average over the above described ensemble of dense set of trajectories. In practice we would generate such an ensemble of $J$ numerical trajectories by the continued application of the edge tracking procedure, recording the extended trajectories similar to those in Fig. \ref{fig:indicator_quantities}.

\subsection{Structural properties}\label{sec:structural}

\begin{figure}
\centerline{\includegraphics[width=\linewidth]{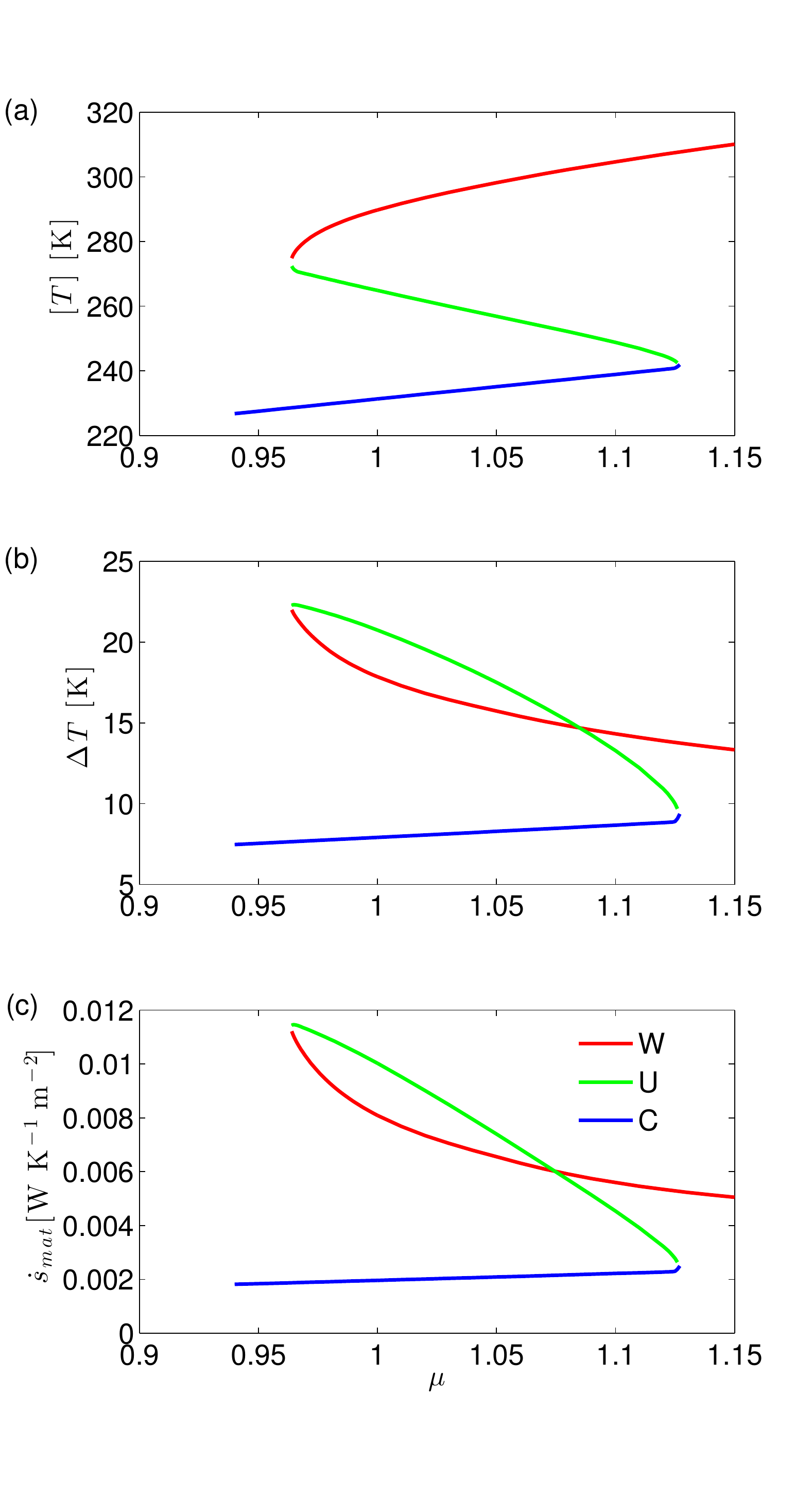}}
\caption{Bifurcation diagrams. The dependence of (a) the average temperature $[T]$ (b) the temperature difference between low and high latitudes $\Delta T$ and (c) the material entropy production $\dot{s}_{mat}$ on the relative solar strength $\mu$ are shown for the three equilibrium solutions of the 1d EBM (\ref{eq:pde}).} \label{fig:bif_diags}
\end{figure}

In Fig. \ref{fig:constitutive_rel} projections of the paths of equilibria onto planes spanned by the different combinations of two of the three variables are also shown. Two of them is reproduced in Fig. \ref{fig:bif_diags} (a) and (b). The first one is in fact a classical result: the bifurcation diagram for the average temperature with the relative solar strength being the bifurcation parameter; {see e.g. Fig. 10.6 of \cite{GC:1987} or Fig. 1 of \cite{npg-8-211-2001}.} A reading of this diagram, {aided for this purpose with annotation by \cite{npg-8-211-2001}}, can be given as follows. 

Starting from the present-day climate, if the solar strength $\mu$ adiabatically decreases, the climate cools (advancing to the left along the red curve), and at a critical point, $\mu_{w\rightarrow c}$, it tips from a relatively warm state to a cold snowball state. (Note that the tipping trajectory is shown in Fig. \ref{fig:constitutive_rel} approximately as the borderline of the surface: the thin blue line belonging to the smallest value of $\mu$ represented there.) If now $\mu$ slowly increases again, the system would not follow the same route back, but the cold climate gradually warms (advancing to the right along the blue curve), upto a point, $\mu_{c\rightarrow w}$, where the basin of attraction of the cold state vanishes, and then the climate tips from a relatively cold snowball state to a warm state. With $\mu$ decreasing again, the present-day climate can be `restored' at the present-day value of $\mu$. That is, changing $\mu$ very slowly in a back-and-forth manner in a range including $\mu_{w\rightarrow c}$ and $\mu_{c\rightarrow w}$, a hysteresis loop is realized. The slowly forced nonautonomous system is governed by the structural properties of the autonomous system.

At the tipping point $\mu_{w\rightarrow c}$ ($\mu_{c\rightarrow w}$) the warm (cold) and the unstable states become identical, and hence the corresponding branches of the bifurcation diagram are connected. In accord with this, the paths of equilibria in the surface $h$, shown in Fig. \ref{fig:constitutive_rel}, are also connected.

\begin{figure}
\centerline{\includegraphics[width=\linewidth]{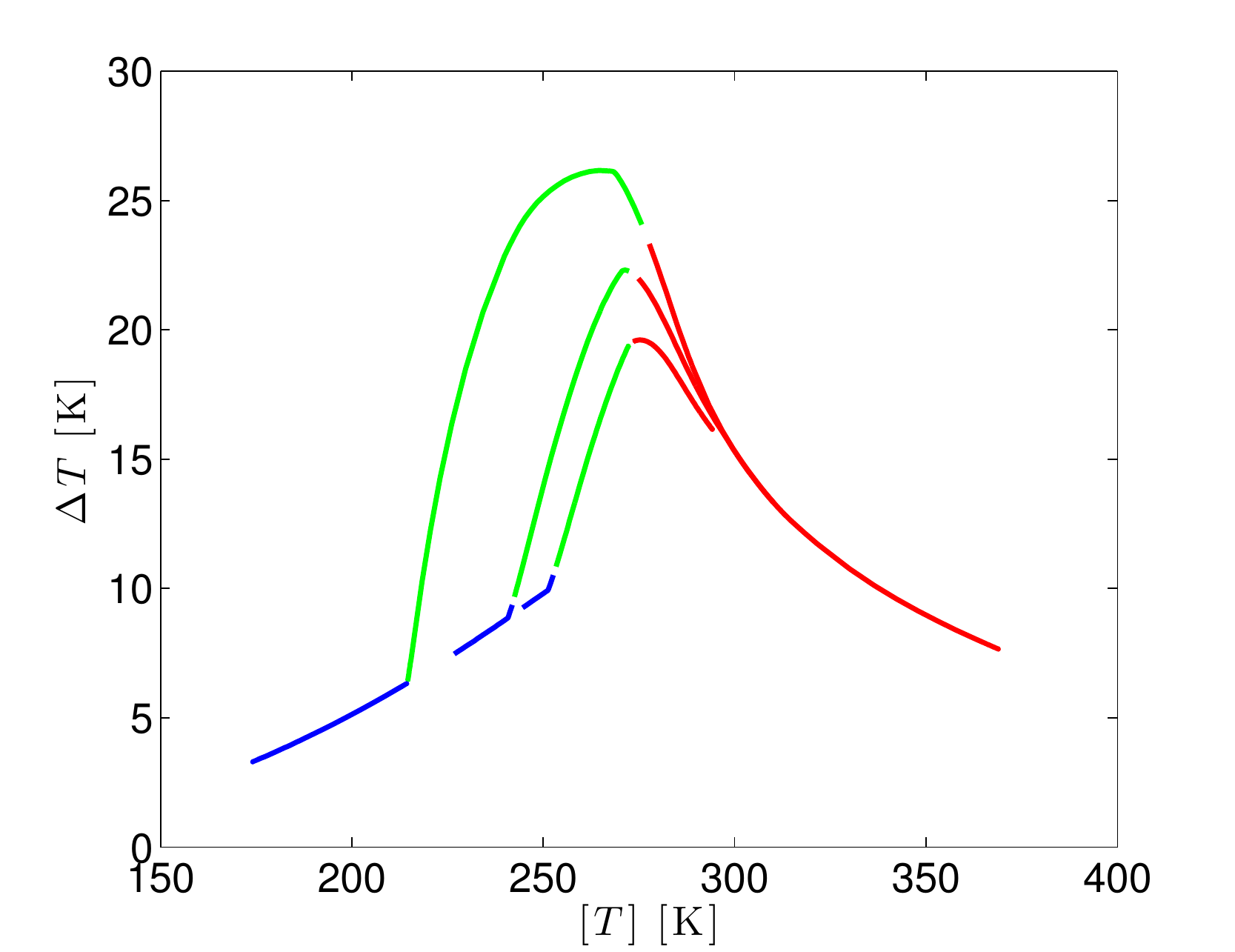}}
\caption{$\Delta T$ plotted against $[T]$ taken from Fig. \ref{fig:bif_diags} (b) and (a), respectively. Three parameter settings are represented: $\alpha_{max} = 0.5$, 0.6, and 0.85, which values in this order result in increasing $[T]$ in the cold state.} \label{fig:DT_vs_aveT}
\end{figure}

Branches of this bifurcation diagram belonging to the three equilibria, $E$, are given by three functions: $[T]=f_E(\mu)$, $E=C,U,W$ for the cold, unstable, and warm states, respectively. The inverse functions of these can be combined into a single function: $\mu=f^{-1}([T])$. In Fig. \ref{fig:constitutive_rel} the projection onto the $\mu$-$\Delta T$ plane results in a similar bifurcation-like diagram, which is reproduced in Fig. \ref{fig:bif_diags} (b). It can be given formally as: $h(f_E(\mu),\Delta T,\mu)=0$, $E=C,U,W$, from which we can construct explicit functions $\Delta T=\delta_E(\mu)$. The third possible projection onto the $[T]$-$\Delta T$ plane is given as: $g([T],\Delta T)=h([T],\Delta T,f^{-1}([T]))=0$, and it is reproduced in Fig. \ref{fig:DT_vs_aveT}.

The use of $\alpha_{max} = 0.6$, as opposed to the original value 0.85 used by 
\cite{Ghil:1976}, results in substantially increased $[T]$ in the cold state and a reduced range of bistability. The quadratic tangency locally e.g. at $\mu_{w\rightarrow c}$ is fairly well-visible in Fig. \ref{fig:bif_diags} (a). From Eq. (\ref{eq:0dEBM}) it can be derived as follows:

\begin{equation}\label{eq:quadratic}
\begin{split}
  \mu=f^{-1}([T]) &\approx \frac{\hat{\sigma}(T_{w\rightarrow c}) T_{w\rightarrow c}^4}{\hat{Q}[1 - \hat{\alpha}(T_{w\rightarrow c})]} \\ &+ \left.\frac{1}{2}\frac{d^2\mu}{d[T]^2}\right|_{T_{w\rightarrow c}}([T]-T_{w\rightarrow c})^2,
\end{split}
\end{equation} 
where $T_{w\rightarrow c}=f_W(\mu_{w\rightarrow c})$. This feature of a saddle-node type bifurcation is a robust one in the climate model hierarchy, as exemplified for an intermediate complexity GCM PlaSim by Fig. 1 of \citep{QJ:QJ543}. Such an observation was made early on by 
\cite{WM:1975}. {We mention that \cite{npg-17-113-2010} explored the dependence of the position $\mu_{w\rightarrow c}$ of the tipping point and its `sharpness' $d^2\mu/d[T]^2|_{T_{w\rightarrow c}}$ on another parameter, which can be related most closely to our (maximal) slope max$[d\hat{\alpha}/d[T]]$. (Strictly speaking the latter is not a parameter, nevertheless, a functional relationship between this and other diagnostic quantities could indeed be constructed.)}

The quadratic tangency of the saddle-node bifurcation of $[T]$ is inherited by $\Delta T$ and $\dot{s}_{mat}$, although it is not so obvious in Fig. \ref{fig:bif_diags} (b) and (c): the respective warm and unstable branches meet in rather sharp points. To see why, first we note that we have a relationship between the slopes through the chain rule of differentiation:

\begin{equation}
  \frac{d\Delta T}{d\mu} = \frac{d\Delta T}{d[T]}\frac{d[T]}{d\mu}.
\end{equation} 
The connection is provided by the function $g([T],\Delta T)=0$. The sign of the slope at the tipping point $d\Delta T/d[T]|_{[T]=T_{w\rightarrow c}}=(dg/d[T])/(dg/d\Delta T)|_{([T],\Delta T)=(T_{w\rightarrow c},\Delta T_{w\rightarrow c})}$, where $g(T_{w\rightarrow c},\Delta T_{w\rightarrow c})=0$, determines whether the `vertical' order of the stable and unstable branches of the diagram of $\Delta T$ with respect to that of $[T]$ flips in the vicinity of the tipping point. When it does flip, $\delta_W(\mu)$ ($\delta_U(\mu)$) becomes a convex (concave) function, while $f_W(\mu)$ ($f_U(\mu)$) remains concave (convex). The bifurcation diagram prompts that $d\Delta T/d[T]|_{[T]=T_{w\rightarrow c}}$ is negative, however, it should be relatively small because of the sharp tipping point. Indeed Fig. \ref{fig:DT_vs_aveT} shows that for $\alpha_{max}=0.6$ the maximum of $\Delta T$ occurs very near the tipping point. In the same figure the cases of $\alpha_{max}=0.5$ and the original 0.85 are also represented, showing that the slopes $d\Delta T/d[T]|_{[T]=T_{w\rightarrow c}}$ have opposite signs, and are not negligible. Accordingly, we observe the opposite vertical order of the $W$ and $U$ branches, and the quadratic tangency in both cases is well visible (not shown, {although for the original $\alpha_{max}=0.85$ qualitatively similar results are shown by Fig. 7 of \cite{Ghil:1976} for a similar quantity $\lim\limits_{t \to \infty} [T(x=0,t)-T(x=1,t)]$}).

We can extend this analysis to the entropy production. Considering the approximation (\ref{eq:2box_smat}), the slope is obtained as:

\begin{equation}
  \begin{split}
    \frac{d\dot{s}_{mat}}{d\mu} \approx \frac{\Delta T}{[T]^2}\Biggl[2\hat{k}([T])\left(\frac{d\Delta T}{d[T]} - \frac{\Delta T}{[T]}\right) \\ + \Delta T\frac{d\hat{k}}{d[T]}\Biggr]\frac{d[T]}{d\mu}.
  \end{split}
\end{equation} 
This reveals that the flipping of the stable and unstable branches, when varying e.g. $\alpha_{max}$, is controlled not only by the sign of the slope $d\Delta T/d[T]|_{[T]=T_{w\rightarrow c}}$, but also by the ratio $\Delta T/[T]$ and by the slope $d\hat{k}/d[T]|_{[T]=T_{w\rightarrow c}}$. However, the latter is negligible here, and also $\Delta T/[T]$ is a relatively small value typically, and so the vertical order of the $W$ and $U$ branches of the diagrams of $\Delta T$ and $\dot{s}_{mat}$ are always the same, except for a very short range of $\alpha_{max}$. The diagrams for $\alpha_{max} = 0.6$ in Fig. \ref{fig:bif_diags} (b) and (c) admit the same vertical order. Furthermore, these diagrams are very similar overall, in the whole range of bistability.

With $\alpha_{max}=0.6$ in a large portion of the range of bistability, for about $\mu<1.07$, the unstable climate is more out of thermodynamic equilibrium than the corresponding stable climates, e.g. $\delta_U(\mu)>\delta_W(\mu)$, which is an interesting contrast to the behavior in terms of the mean temperature, namely, $f_U(\mu)<f_W(\mu)$ for any $\mu\in[\mu_{w\rightarrow c},\mu_{c\rightarrow w}]$. With $\alpha_{max}=0.5$ the stable warm climate is more out of thermodynamic equilibrium than the unstable climate for any $\mu$ (not shown), $\delta_U(\mu)<\delta_W(\mu)$, which is already similar to the unchanged relation $f_U(\mu)<f_W(\mu)$. It is unchanged since $[T]$ is still governed by a well-defined 0d EBM. However, from the point of view of the transient dynamics, $\Delta T$ does not behave similarly to $[T]$ even with $\alpha_{max}=0.5$, since during the transient much larger thermodynamic disequilibrium is possible than in the stable warm state, similarly as with $\alpha_{max}=0.6$ shown by Fig. \ref{fig:constitutive_rel}. That is, in neither of these two cases can a 0d EBM for $\Delta T$ be well-defined. This feature is a robust one in a large range of $\alpha_{max}$, which includes the original value 0.85. However, the position of the (green) path of the unstable solutions in the surface $h$ varies considerably with $\alpha_{max}$. Accordingly, a transient increase of $\Delta T$ occurs sometimes towards not the warm but the cold state. For some values of $\alpha_{max}$ there exists a critical value of $\mu$ where the unstable path crosses the ridge of $h$, and so no transient increase will occur in either direction. For the original $\alpha_{max}=0.85$ this value happens to be very close to the present day value of $\mu=1$. Consequently, the warm-to-cold B-tipping at $\mu_{w\rightarrow c}<1$ would start with a transient increase of $\Delta T$ in this model.

\section{Summary and discussion}\label{sec:conclusions}

In this paper we applied the so-called edge tracking technique to find the unstable solution of a geophysical problem, namely, that of a diffusive 1d energy balance climate model (EBM), {the Ghil-Sellers model}. The original PIM-triple algorithm is applicable generally in order to construct long trajectories on a nonattracting dynamical object, e.g., a chaotic saddle. The edge tracking algorithm is a more efficient version of it involving an iterative bisection procedure, which can be used when the unstable solution is due to bistability in the system, or, when a scalar quantity can indicate the type of flow regime that the trajectory is exploring temporarily. The unstable solution in the latter more general situation -- separating different regimes -- is referred to as an `edge state'. The edge tracking algorithm was proposed and successfully applied to various shear flow problems by Eckhardt and co-workers. It has been applied now for the first time to a geophysical problem featuring bistability. The unstable solution of the 1d EBM can be found also by a boundary value problem solver algorithm. This possibility provides us a reference solution, against which the solution obtained by edge tracking can be compared. In the examined cases we found excellent agreement.  

We examined the influence of tolerance parameters of the algorithm on its robustness, efficiency, and the error of approximation. We find, for example, that a too stringent tolerance of bracketing the stable manifold of the unstable state would make the procedure inefficient, and in fact taking a single bisection, i.e., avoiding iteration completely, would be most efficient in our case -- and in most practical cases, presumably. Furthermore, we find that in the strongly diffusive 1d EBM the convergence to the edge state is very fast, and so the approximation does not improve already from the second cycle of the iterative/cyclic edge tracking procedure (assuming reasonable tolerances imposed).

A related effect to the fast convergence is the rapid approach of the heteroclinic trajectory, which connects the unstable state (saddle fixed point) with a stable state (node fixed point), by arbitrarily initialized trajectories, whereby they would approach the stable state subsequently evolving slower and closely confined to the heteroclinic trajectory. The unique heteroclininc trajectory, which is of dimension one, thus, dictates a functional relationship between any pair of observables (prognostic or diagnostic) after a short transient time. We called such a relationship between the average temperature $[T]$ and the temperature difference $\Delta T$ between high and low latitudes a climate constitutive relationship. The difference $\Delta T$ is a simple measure of thermodynamic disequilibrium, the characterization of which is another main goal of the present analysis. We found that $\Delta T$ has a single maximum as a function of $[T]$ at about 270 K, which is approximately unchanged within the whole range of bistability with respect to the solar strength $\mu$. The investigation of this interesting behavior is a subject of ongoing research. It has to do with the fact that at about this temperature water freezes, with which the albedo changes abruptly, and since it is the average temperature concerned, the maximal transport occurs when about half of the planet is snow-covered (averaging over a year). As $[T_U]$ of the unstable state, in contrast, depends considerably on $\mu$, a transient increase and nonmonotonic evolution of $\Delta T$ towards a stable state -- unlike the evolution of $[T]$ -- is a typical behavior.

We constructed bifurcation diagrams, representing structural properties of the system, in terms of $[T]$ and, as new result, $\Delta T$ and a related quantity $\dot{s}_{mat}$, the material entropy production in the course of meridional heat transport -- another measure of thermodynamic disequilibrium. In these diagrams, beside the branches of the two stable solutions, the third branch of the unstable solutions, which connects the two other branches at the tipping points, is also present. At the warm-to-cold tipping point we observe a quadratic tangency of the warm ($W$) and unstable ($U$) branches of $[T_{W/U}(\mu)]$, and this is inherited by those of $\Delta T_{W/U}(\mu)$ and $\dot{s}_{mat,W/U}(\mu)$ for realistic values of the maximal albedo $\alpha_{max}$ of snow. However, the vertical order of $\Delta T_{W}(\mu)$ and $\Delta T_{U}(\mu)$ is reversed relative to that of $[T]_{W}(\mu)$ and $[T]_{U}(\mu)$ at the warm-to-cold tipping point for large enough realistic values of $\alpha_{max}$. This applies also to $\dot{s}_{mat}$. Accordingly, $\Delta T_{W}(\mu)$ is a convex and decreasing function near the tipping point. This feature, characterized also by a negative slope $d\Delta T/d[T]$ seen in Fig. \ref{fig:DT_vs_aveT}, reflects the effect of polar amplification, first studied by \cite{TUS:TUS466} and \cite{Sellers:1969} in EBMs, and by \cite{WM:1975} in an intermediate complexity GCM. {We note that as a result of some parameter change, e.g. $\alpha_{max}$ decreased to 0.5, the path that crosses the ridge of the surface $h$ shown in Fig. \ref{fig:constitutive_rel} -- in the spirit of the discussion at the end of Sec. \ref{sec:structural} -- can be that of the stable warm states (instead of the unstable ones). Consequently, $\Delta T_{W}(\mu)$ becomes nonmonotonic, implying that between the tipping point and the value of $\mu$ where the maximum of $\Delta T_{W}(\mu)$ occurs, there is no polar amplification, but it is in fact reversed.}

\begin{figure}
    \begin{center}
        \includegraphics[width=\linewidth]{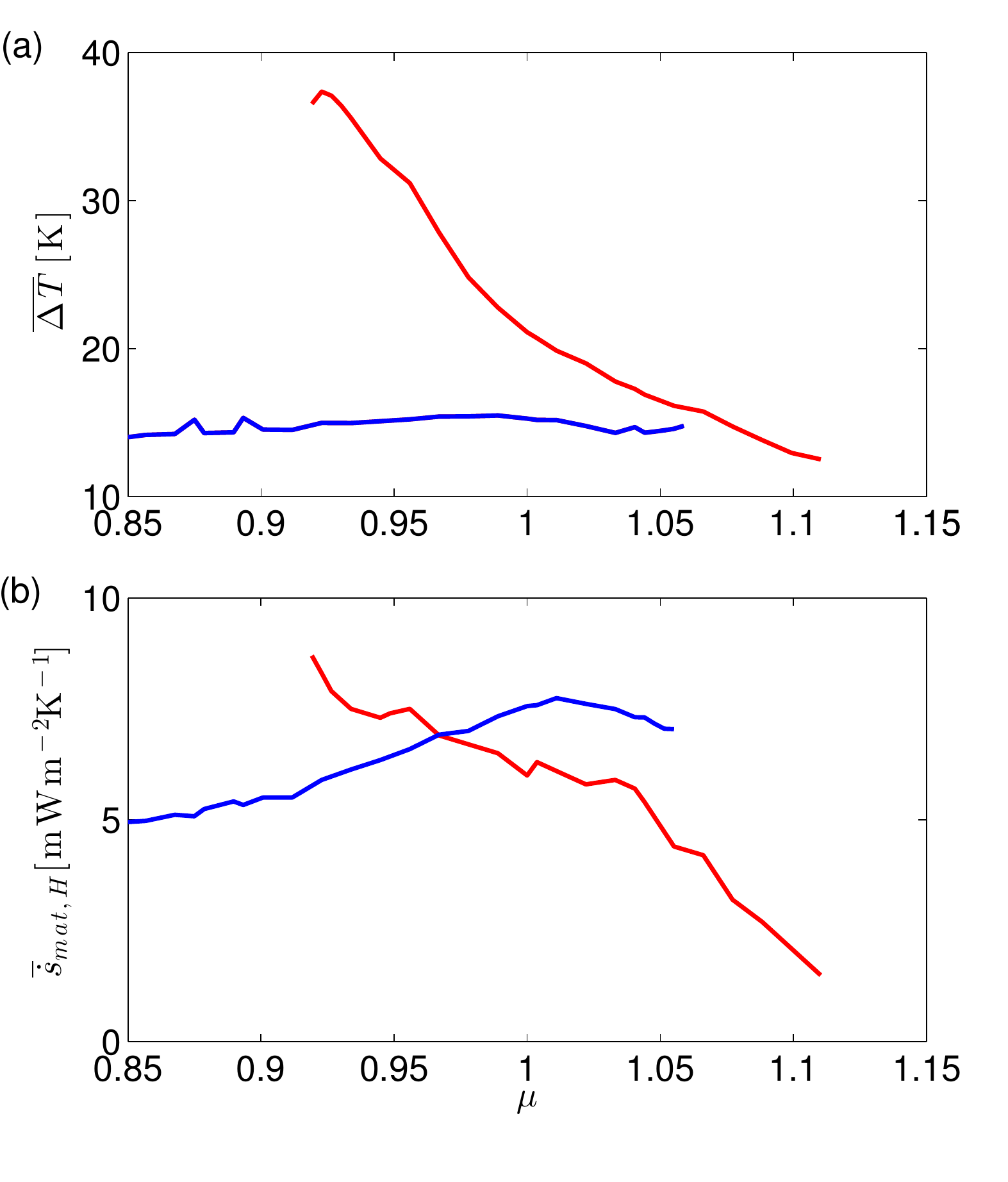}
        \caption{\label{fig:Plasim_noneq_thermodyn} Thermodynamic disequilibrium properties in PlaSim. (a) Long-time average temperature difference $\overline{\Delta T}$ between similar boxes to those defined for the 1d EBM, based on extrapolated temperature values on the 1000 hPa (or sea surface) level. (b) Long-time average entropy production in the course of horizontal ($H$) processes. The model configuration was adapted from \cite{Boschi2013}, and a fixed CO$_2$ concentration of 360 ppm was considered.
	}
    \end{center}
\end{figure}

It is interesting to observe that the simple 1d EBM considered in this study captures, at least qualitatively, some of the features of the much more complex PlaSim model. As noted earlier, the dependence of $[T]$ on $\mu$ in the cold and warm states is similar to what was reported for PlaSim [compare Figs. 1 of \citep{QJ:QJ543} and \ref{fig:bif_diags} (a)]. Figure \ref{fig:Plasim_noneq_thermodyn} (a) shows the long-term average of the $\mu$-dependence of the difference $\overline{\Delta T}$ between the sea surface temperatures of the high and low latitudes of PlaSim, which has a good degree of correspondence with what is shown in Fig. \ref{fig:bif_diags} (b). Considering the entropy production $\dot{s}_{mat}$ in Fig. \ref{fig:bif_diags} (c), on the other hand, one may be confused by the fact that it decreases with $\mu$ in the warm state, whereas the opposite was reported for PlaSim [Fig. 4 (a) of \citep{QJ:QJ543}]. The EBM used here is not able to capture vertical processes, which, as discussed by \cite{PGATL:2012}, give the strongest contribution to the entropy production. If now one computes in PlaSim, following \cite{LFR:2011}, the contribution to the entropy production due to large scale horizontal processes, one obtains a bifurcation diagram [Fig. \ref{fig:Plasim_noneq_thermodyn} (b)] which has again qualitative similarities with the one in Fig. \ref{fig:bif_diags} (c).

As the convexity properties of the bifurcations diagrams for PlaSim in Fig. \ref{fig:Plasim_noneq_thermodyn} (a) and (b) are hard to be determined, the vertical order of their $W$ and $U$ branches is not really possible to guess even in the vicinity of the tipping point. Our objective for future research is to construct these diagrams complete with the unstable $U$ branches, applying the edge tracking technique. The relative dissimilarity of these diagrams (showing at least the stable branches), in comparison with the similarity of the corresponding diagrams of the EBM in Fig. \ref{fig:bif_diags} (b) and (c), is another sign of the relative complexity of PlaSim over the 1d EBM.

A further limitation of the EBM is that the thermodynamic efficiency \citep{PhysRevE.80.021118}, another measure of nonequilibrium thermodynamics, cannot be defined in its terms, since the fluid dynamics is eliminated from the model, or in other words, it is not explicitly represented. In PlaSim, however, the thermodynamic efficiency can be evaluated, and it was found to increase before both the warm-to-cold and the cold-to-warm tipping (see Fig. 3 (a) of \citep{QJ:QJ543}). Since the unstable branch connects the stable branches at the tipping points, it is reasonable to think, then, that the atmosphere under the unstable climate is a more efficient thermal engine than under either the warm or the cold stable climates, everywhere in the range of bistability. Whether this is the case is yet to be seen, once the unstable climate states are successfully constructed by the edge tracking technique. Then, it would be also an interesting question whether a maximal efficiency is achieved in a nonequilibrium dynamical state, like $\Delta T$ of the 1d EBM, or, whether a constitutive relationship between the efficiency and some global average temperature has more complex characteristics than just a single maximum. The following observation can also be made regarding the bifurcation diagram shown in Fig. 3 (a) of \citep{QJ:QJ543}. At either of the tipping points locally no quadratic tangency of the stable branches can be seen -- unlike the situation with the total entropy production shown in Fig. 4 (a) of \citep{QJ:QJ543}. This makes any guess about the unstable branch harder, and thereby further motivates the search for the unstable states in PlaSim. 

{Another line of our interests for future work is concerned with the response \citep{Lucarini:2009,npg-18-7-2011} of the climate system near a tipping point. We will investigate nonlinear terms of the response, expected to be influenced substantially by the nearby unstable state.}

Before attempting to construct the edge states by applying the edge tracking technique for a full climate model, in an ongoing work we try to do just that for a bistable Earth-like but dry model atmosphere realized by the software suite the `Portable University Model of the Atmosphere' (PUMA), which constitutes the dynamical core of PlaSim. We represent the negative feedback mechanism that creates bistability by a phenomenological model of the ice-albedo feedback, specifying the surface albedo by a suitable function of the surface temperature. Following that we will do the same with PlaSim. Possible challenges with that include the interpolation between supersaturated and unsaturated moist states when bisecting. 

Earth-like planets of different orbital parameters, in particular: the ratio of the lengths of the day and the year, exhibit substantially different multistability properties~\citep{ASNA:ASNA201311903,Boschi2013}. It seems to be of great interest, in perspective, to investigate such structural changes by using the edge tracking method.

Here, by finding and characterizing unstable solutions, we wish also to initiate work on the understanding how global instability determines the variability of nonautonomous complex systems like climate.

\begin{acknowledgements}
The authors would like to thank Tobias Kreilos and Salvatore Pascale for stimulating exchanges, {and Tam\'as T\'el for reading our manuscript}. {Numerous constructive comments by Michael Ghil to our manuscript submission helped greatly to improve the quality of our work; and we thank another anonymous reviewer for calling our attention to the relevant publication by~\cite{DW:2005}.} The presented work is part of the project NAMASTE -- Thermodynamics of the Climate System, which is supported financially by the European Research Council, under grant agreement No. 257106. {Support of the Excellence Cluster `Integrated Climate System Analysis and Prediction' (CliSAP) at the University of Hamburg is also acknowledged.}
\end{acknowledgements}

\appendix

\section{Phase III of the edge tracking procedure}\label{apdx:phase_III}

After advancing the profiles in phase II of the edge tracking procedure (described in Sec. \ref{sec:edgetracking}) it might be the case that if we restart the simulation and further advance the profiles (were we to proceed with phase I), both of them end up with the same climate. This can happen if an adaptive time step (and/or implicit) integrator is used, such as e.g. Matlab's \texttt{pdepe}, which initiates the integration with an algorithm other than that is used for the rest of the integration procedure. In the vicinity of the edge or its stable manifold, trajectories are very sensitive to perturbations regarding the outcome. Changing the integrator scheme at arbitrary times (extrinsic to the treated dynamical system) is a kind of numerical perturbation, which is enough to change the outcome when the simulation is interrupted and restarted. This can be expressed in a way that the numerical evolution operator is a nonautonomous two-time operator: $\mathcal{E}^*_{t_1,t_2}[\cdot]$, $t_1<t_2$. Given certain resolutions in space and time for the numerical solution, a too small choice for $\varepsilon_2$ will result in the described problem. 

If for example $\mathcal{E}^*_{t_{j+1},t}[\mathcal{E}^*_{t_j,t_{j+1}}[T_{c,I,j}]]$ ends up with the warm climate, then it is reinitialized with the subtraction of a small constant number, $T_{c,0,j+1}=\mathcal{E}^*_{t_j,t_{j+1}}[T_{c,I,j}] - \varepsilon_3$, and we check if it fixes the problem, i.e., if $\mathcal{E}^*_{t_{j+1},t}[\mathcal{E}^*_{t_j,t_{j+1}}[T_{c,I,j}]- \varepsilon_3] \longrightarrow T_C$ for $t\longrightarrow\infty$. If not, then the incremental negative correction is applied again, $T_{c,0,j+1}=\mathcal{E}^*_{t_j,t_{j+1}}[T_{c,I,j}] - 2\varepsilon_3$, and it is repeated until the problem is fixed. A robust choice can be $\varepsilon_3 = \varepsilon_2$, but it may be different from the computationally most efficient choice. 

We note that as an alternative strategy, the error tolerance of the numerical integrator can be set stringently enough, depending on $\varepsilon_2$, that the original problem would not present itself at all. It is yet to be seen which strategy is more efficient, provided that the same accuracy is achieved.

\section{Phase portrait}\label{apdx:phase_portrait}

\begin{figure*}
    \begin{center}
        \includegraphics[width=\linewidth]{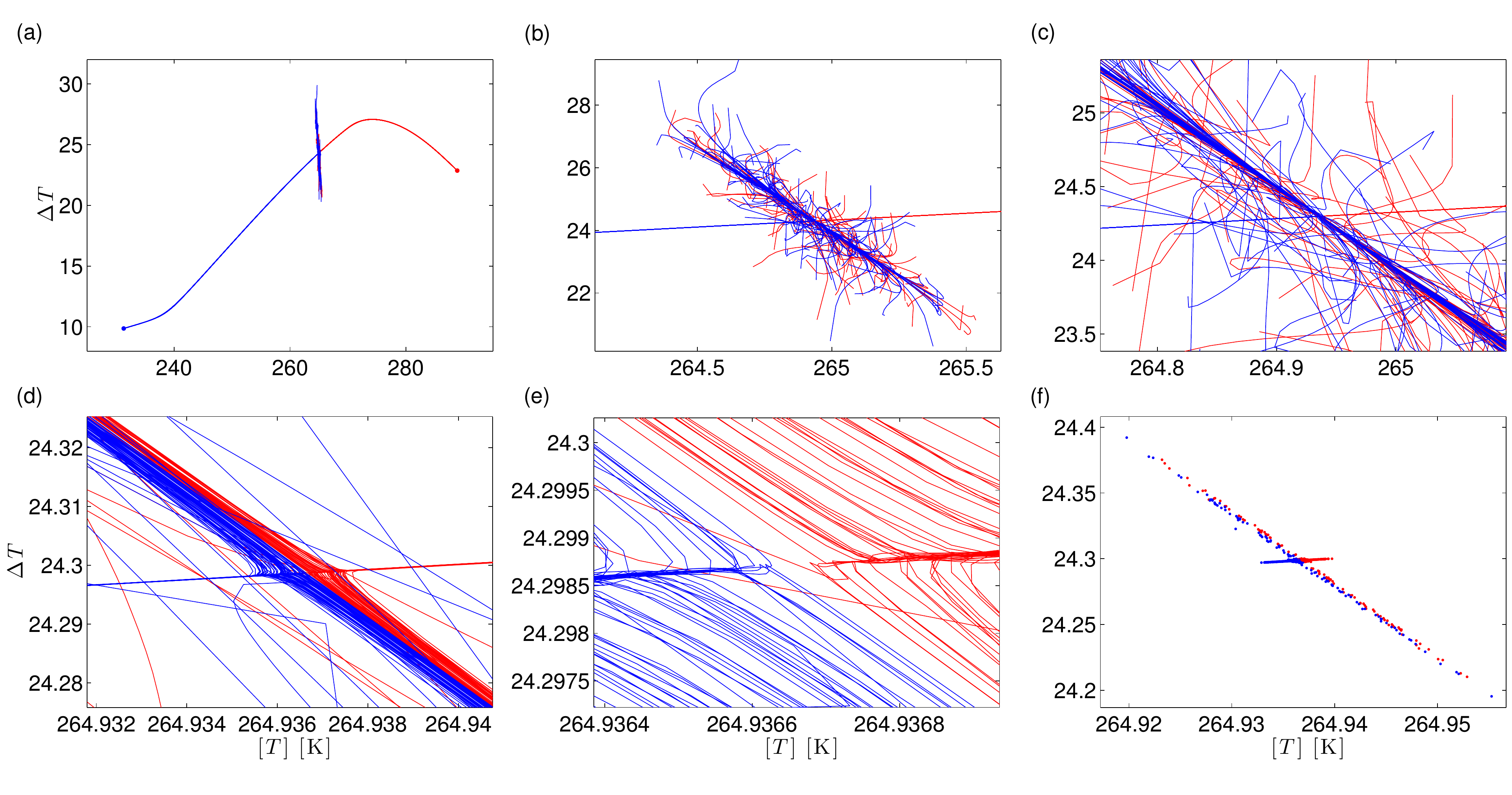}
        \caption{\label{fig:phase_portrait} Phase portrait of the 1d EBM ($\mu=1$, $\alpha_{max}=0.6$) projected onto the $[T]$-$\Delta T$ plane. The phase portrait (a) shows the stable fixed point solutions highlighted by markers in the end of the respective branches of the heteroclinic orbit; at the intersection point of the latter with the nearly vertical basin boundary situated is the unstable (saddle-type) fixed point. A series of zooms (b)-(e) around this saddle point is provided to give further insight explained in the text. The (ratio of the) time scales belonging to the stable and unstable manifolds of the saddle point is indicated by two snapshots of the used trajectories taken the same time (chosen to be 2.352 $\times10^8$ sec) before and after the time instant when the ensemble assumes the smallest size (at 2.364$\times10^8$ sec) (f).
	}
    \end{center}
\end{figure*}

{

The {\em phase portrait} of a dynamical systems is usually represented by a collection of trajectories that sample the various different regimes (types of trajectories) featured by the system. Here we wish to supplement the discussion in the core text by a visual display of some of the concepts mentioned there. Figure \ref{fig:phase_portrait} (a) shows the `skeleton' of the phase portrait of the 1d EBM projected onto the 2D $[T]$-$\Delta T$ plane. For this we use the longest-lived 200 trajectories out of an ensemble of $N=10^6$ trajectories, which were initialized by a random perturbation of the stationary unstable temperature profile: $T_n(x) = T_U(x) + \delta\xi_n(x)$, $n=1,\dots,N$, where, for each $n$, $\xi_n(x)$ may be a uniformly distributed white noise process of zero mean and unit variance, and $\delta$ is the perturbation strength, set to be $\delta=15$ for this exercise. We note that in our numerics for the different ($l$) gridpoints uncorrelated random numbers $\xi_{n,l}$ were generated, with which $T_{n,l} = T_{U,l} + \delta\xi_{n,l}$. As part of the initialization of the numerical integration, derivatives of the profile in the gridpoints are obtained by Matlab's \texttt{pdeval}, which, just like \texttt{pdepe}, assumes a second order approximation of the solution in the gridpoints. That is, the perturbed initial conditions are in fact assumed to be continuous and smooth, unlike white noise. 

Since these are the exponentially few longest-lived trajectories used, we know that initially they stayed very near the {\em basin boundary}, and so the initial ensemble gives a numerical representation of the basin boundary not far from the {\em unstable saddle-type fixed point}. Panel (a) shows that it is a steep, nearly vertical line -- at least in the explored region not far from the saddle point -- which means that it is dominantly $[T]$ that determines which {\em stable node-type fixed point} the system ends up with upon perturbing the unstable solution. (This property is epitomized by 0d EBMs.) The steep line of the basin boundary, being one and the same object as the {\em stable manifold} of the saddle point, aligns with the second {\em eigenvector} of the 2-DOF model, which latter is the same -- with appropriate parametrization of the 2-DOF model -- as the second eigenvector of the 1d EBM projected onto the plane. However, as the initial ensemble is generated by noise perturbations, its members populate the infinite-dimensional {\em phase space} (which is limited in the numerics by the coarsegraining). Therefore, the basin boundary is also infinite-dimensional, and so its projection onto the 2D $[T]$-$\Delta T$ plane is actually area-filling. This is seen in the magnified views in panels (b)-(c). Further magnification in panels (d)-(e) reveals that during their relatively long lifetime the trajectories `become low-dimensional' (d), whereby trajectories eventually going to e.g. the cold climate line up on the cold side of the second eigenvector. (In the course of becoming low-dimensional trajectories also become less `curly'.) Clearly, the closer the trajectory to the eigenvector, the longer it lives; and the longest-lived cold- and warm-side trajectories bracket the saddle point the tightest, where they turn very sharply (e). 

One can follow the ensemble of long-lived trajectories in time. They approach the saddle point moving along its stable manifold, along a relatively straight line in our case. The ensemble assumes its smallest size at about the time when each trajectory approaches the (sole) saddle point most closely in the course of its evolution. (When the saddle set is a fractal set, the time of closest approach is indicated by an emerging clear double-fractal characteristic of the ensemble -- not by its small size.) By this time the trajectories are necessarily close also to the {\em unstable manifold} of the saddle point, which is identical with the (sole) {\em heteroclininc orbit} in our case, and hereafter they move on, towards the stable fixed points, along this unstable manifold. Therefore, snapshots of the ensemble at-, before-, and after the closest approach can represent the saddle point, and its stable and unstable manifolds, respectively. This is referred to as Sprinkler's method~\citep{TG:2006}. These snapshots are shown in panel (f). In addition to the directions of the eigenvectors giving the inclinations of the manifolds at the saddle point, we can extract information also about the eigenvalues. The spread of the first snapshot is about 5 times that of the third snapshot, taken at times equally separated from the second snapshot (not visible because of the relatively small size of the ensemble), which ratio agrees with that of the eigenvalues provided in Sec. \ref{sec:results_edge_tracking}.

} 

\bibliographystyle{spbasic}      

\begin{thebibliography}{54}
\providecommand{\natexlab}[1]{#1}
\providecommand{\url}[1]{{#1}}
\providecommand{\urlprefix}{URL }
\expandafter\ifx\csname urlstyle\endcsname\relax
  \providecommand{\doi}[1]{DOI~\discretionary{}{}{}#1}\else
  \providecommand{\doi}{DOI~\discretionary{}{}{}\begingroup
  \urlstyle{rm}\Url}\fi
\providecommand{\eprint}[2][]{\url{#2}}

\bibitem[{Ashwin et~al(2012)Ashwin, Wieczorek, Vitolo, and Cox}]{AWVC:2012}
Ashwin P, Wieczorek S, Vitolo R, Cox P (2012) Tipping points in open systems:
  bifurcation, noise-induced and rate-dependent examples in the climate system.
  Phil Trans R Soc A 371(1962):1166--1184

\bibitem[{Berry et~al(1945)Berry, Bollay, and Beers}]{BBB:1945}
Berry FAJ, Bollay E, Beers NRe (1945) Handbook of Meteorology. McGraw-Hill

\bibitem[{B\'odai et~al(2011)B\'odai, K\'arolyi, and T\'el}]{BKT:2011}
B\'odai T, K\'arolyi G, T\'el T (2011) Fractal snapshot components in chaos
  induced by strong noise. Phys Rev E 83:046,201,
  \doi{10.1103/PhysRevE.83.046201},
  \urlprefix\url{http://link.aps.org/doi/10.1103/PhysRevE.83.046201}

\bibitem[{B\'odai et~al(2013)B\'odai, Altmann, and Endler}]{PhysRevE.87.042902}
B\'odai T, Altmann EG, Endler A (2013) Stochastic perturbations in open chaotic
  systems: Random versus noisy maps. Phys Rev E 87:042,902,
  \doi{10.1103/PhysRevE.87.042902},
  \urlprefix\url{http://link.aps.org/doi/10.1103/PhysRevE.87.042902}

\bibitem[{Bordi et~al(2013)Bordi, Fraedrich, Sutera, and Zhu}]{BFSZ:2013}
Bordi I, Fraedrich K, Sutera A, Zhu X (2013) On the effect of decreasing
  $\textrm{CO}_2$ concentration in the atmosphere. Climate Dynamics
  40(3-4):651--662, \doi{10.1007/s00382-012-1581-z},
  \urlprefix\url{http://dx.doi.org/10.1007/s00382-012-1581-z}

\bibitem[{Boschi et~al(2013)Boschi, Lucarini, and Pascale}]{Boschi2013}
Boschi R, Lucarini V, Pascale S (2013) Bistability of the climate around the
  habitable zone: A thermodynamic investigation. Icarus 226(2):1724 -- 1742,
  \doi{http://dx.doi.org/10.1016/j.icarus.2013.03.017},
  \urlprefix\url{http://www.sciencedirect.com/science/article/pii/S0019103513001309}

\bibitem[{Budyko(1969)}]{TUS:TUS466}
Budyko MI (1969) The effect of solar radiation variations on the climate of the
  earth. Tellus 21(5):611--619, \doi{10.1111/j.2153-3490.1969.tb00466.x},
  \urlprefix\url{http://dx.doi.org/10.1111/j.2153-3490.1969.tb00466.x}

\bibitem[{Dakos et~al(2008)Dakos, Scheffer, van Nes, Brovkin, Petoukhov, and
  Held}]{Dakos23092008}
Dakos V, Scheffer M, van Nes EH, Brovkin V, Petoukhov V, Held H (2008) Slowing
  down as an early warning signal for abrupt climate change. Proceedings of the
  National Academy of Sciences 105(38):14,308--14,312,
  \doi{10.1073/pnas.0802430105},
  \urlprefix\url{http://www.pnas.org/content/105/38/14308.abstract},
  \eprint{http://www.pnas.org/content/105/38/14308.full.pdf+html}

\bibitem[{Dijkstra(2005)}]{Dijkstra:2005}
Dijkstra HA (2005) Nonlinear Physical Oceanography. Springer, Dordrecht

\bibitem[{Dijkstra and Weijer(2005)}]{DW:2005}
Dijkstra HA, Weijer W (2005) Stability of the global ocean circulation: Basic
  bifurcation diagrams. Journal of Physical Oceanography 35(6):933--948,
  \doi{10.1175/JPO2726.1}, \urlprefix\url{http://dx.doi.org/10.1175/JPO2726.1}

\bibitem[{Dijkstra et~al(2014)Dijkstra, Wubs, Cliffe, Doedel, Dragomirescu,
  Eckhardt, Gelfgat, Hazel, Lucarini, Salinger, Phipps, Sanchez-Umbria,
  Schuttelaars, Tuckerman, and Thiele}]{Num_bif_meth}
Dijkstra HA, Wubs FW, Cliffe AK, Doedel E, Dragomirescu IF, Eckhardt B, Gelfgat
  AY, Hazel AL, Lucarini V, Salinger AG, Phipps ET, Sanchez-Umbria J,
  Schuttelaars H, Tuckerman LS, Thiele U (2014) Numerical bifurcation methods
  and their application to fluid dynamics: analysis beyond simulation. Commun
  Comput Phys 15:1--45

\bibitem[{Ditlevsen and Johnsen(2010)}]{GRL:GRL27358}
Ditlevsen PD, Johnsen SJ (2010) Tipping points: Early warning and wishful
  thinking. Geophysical Research Letters 37(19):n/a--n/a,
  \doi{10.1029/2010GL044486},
  \urlprefix\url{http://dx.doi.org/10.1029/2010GL044486}

\bibitem[{Donnadieu et~al(2004)Donnadieu, Ramstein, Fluteau, Roche, and
  Ganopolski}]{DRFRG:2004}
Donnadieu Y, Ramstein G, Fluteau F, Roche D, Ganopolski A (2004) The impact of
  atmospheric and oceanic heat transports on the sea-ice-albedo instability
  during the neoproterozoic. Climate Dynamics 22(2-3):293--306,
  \doi{10.1007/s00382-003-0378-5},
  \urlprefix\url{http://dx.doi.org/10.1007/s00382-003-0378-5}

\bibitem[{Dwyer and Pettersen(1973)}]{DP:1973}
Dwyer HA, Pettersen (1973) Time-dependent global energy modeling. J Appl Meteor
  12:36--42

\bibitem[{Faranda et~al(2012)Faranda, Lucarini, Manneville, and
  Wouters}]{FLMW:2012}
Faranda D, Lucarini V, Manneville P, Wouters J (2012) On using extreme values
  to detect global stability thresholds in multi-stable systems: The case of
  transitional plane {Couette} flow. arXiv:12110510 [mathDS]

\bibitem[{Fraedrich(2012)}]{Fraedrich:2012}
Fraedrich K (2012) A suite of user-friendly global climate models: Hysteresis
  experiments. The European Physical Journal Plus 127(5):1--9,
  \doi{10.1140/epjp/i2012-12053-7},
  \urlprefix\url{http://dx.doi.org/10.1140/epjp/i2012-12053-7}

\bibitem[{Freidlin and Wentzell(1984)}]{FW:1984}
Freidlin MI, Wentzell AD (1984) Random Perturbations of Dynamical Systems.
  Springer, New York

\bibitem[{Ghil(1976)}]{Ghil:1976}
Ghil M (1976) Climate stability for a {Sellers}-type model. J Atmos Sci
  33:3--20

\bibitem[{Ghil(2001)}]{npg-8-211-2001}
Ghil M (2001) Hilbert problems for the geosciences in the 21st century.
  Nonlinear Processes in Geophysics 8(4/5):211--211,
  \doi{10.5194/npg-8-211-2001},
  \urlprefix\url{http://www.nonlin-processes-geophys.net/8/211/2001/}

\bibitem[{Ghil and Childress(1987)}]{GC:1987}
Ghil M, Childress S (1987) Topics in Geophysical Fluid Dynamics: Atmospheric
  Dynamics, Dynamo Theory, and Climate Dynamics. Springer-Verlag, New York

\bibitem[{Grassl(1981)}]{Grassl:1981}
Grassl H (1981) The climate at maximum entropy production by meridional
  atmospheric and oceanic heat fluxes. Quarterly Journal of the Royal
  Meteorological Society 107(451):153--166, \doi{10.1002/qj.49710745110},
  \urlprefix\url{http://dx.doi.org/10.1002/qj.49710745110}

\bibitem[{de~Groot and Mazur(1969)}]{GM:1969}
de~Groot SR, Mazur P (1969) Non-equilibrium thermodynamics. North-Holland
  Publishing Company, Amsterdam-London

\bibitem[{Hoffman et~al(1998)Hoffman, Kaufman, Halverson, and
  Schrag}]{Hoffman28081998}
Hoffman PF, Kaufman AJ, Halverson GP, Schrag DP (1998) A {Neoproterozoic
  Snowball Earth}. Science 281(5381):1342--1346,
  \doi{10.1126/science.281.5381.1342},
  \urlprefix\url{http://www.sciencemag.org/content/281/5381/1342.abstract},
  \eprint{http://www.sciencemag.org/content/281/5381/1342.full.pdf}

\bibitem[{Iansiti et~al(1985)Iansiti, Hu, Westervelt, and
  Tinkham}]{PhysRevLett.55.746}
Iansiti M, Hu Q, Westervelt RM, Tinkham M (1985) Noise and chaos in a fractal
  basin boundary regime of a josephson junction. Phys Rev Lett 55:746--749,
  \doi{10.1103/PhysRevLett.55.746},
  \urlprefix\url{http://link.aps.org/doi/10.1103/PhysRevLett.55.746}

\bibitem[{Jabri(2003)}]{Jabri:2003}
Jabri Y (2003) The Mountain Pass Theorem, Variants, Generalizations and Some
  Applications. Cambridge University Press

\bibitem[{Lai and T\'el(2011)}]{LT:2011}
Lai YC, T\'el T (2011) Transient Chaos. Springer, New York

\bibitem[{Lenton et~al(2008)Lenton, Held, Kriegler, Hall, Lucht, Rahmstorf, and
  Schellnhuber}]{Lenton12022008}
Lenton TM, Held H, Kriegler E, Hall JW, Lucht W, Rahmstorf S, Schellnhuber HJ
  (2008) Tipping elements in the {Earth}'s climate system. Proceedings of the
  National Academy of Sciences 105(6):1786--1793,
  \doi{10.1073/pnas.0705414105},
  \urlprefix\url{http://www.pnas.org/content/105/6/1786.abstract},
  \eprint{http://www.pnas.org/content/105/6/1786.full.pdf+html}

\bibitem[{Lucarini(2009{\natexlab{a}})}]{Lucarini:2009}
Lucarini V (2009{\natexlab{a}}) Evidence of dispersion relations for the
  nonlinear response of the lorenz 63 system. Journal of Statistical Physics
  134(2):381--400, \doi{10.1007/s10955-008-9675-z},
  \urlprefix\url{http://dx.doi.org/10.1007/s10955-008-9675-z}

\bibitem[{Lucarini(2009{\natexlab{b}})}]{PhysRevE.80.021118}
Lucarini V (2009{\natexlab{b}}) Thermodynamic efficiency and entropy production
  in the climate system. Phys Rev E 80:021,118,
  \doi{10.1103/PhysRevE.80.021118},
  \urlprefix\url{http://link.aps.org/doi/10.1103/PhysRevE.80.021118}

\bibitem[{Lucarini and Sarno(2011)}]{npg-18-7-2011}
Lucarini V, Sarno S (2011) A statistical mechanical approach for the
  computation of the climatic response to general forcings. Nonlinear Processes
  in Geophysics 18(1):7--28, \doi{10.5194/npg-18-7-2011},
  \urlprefix\url{http://www.nonlin-processes-geophys.net/18/7/2011/}

\bibitem[{Lucarini et~al(2010)Lucarini, Fraedrich, and Lunkeit}]{QJ:QJ543}
Lucarini V, Fraedrich K, Lunkeit F (2010) Thermodynamic analysis of snowball
  {Earth} hysteresis experiment: Efficiency, entropy production and
  irreversibility. Quarterly Journal of the Royal Meteorological Society
  136(646):2--11, \doi{10.1002/qj.543},
  \urlprefix\url{http://dx.doi.org/10.1002/qj.543}

\bibitem[{Lucarini et~al(2011)Lucarini, Fraedrich, and Ragone}]{LFR:2011}
Lucarini V, Fraedrich K, Ragone F (2011) New results on the thermodynamic
  properties of the climate system. J Atmos Sci 68:2438--2458

\bibitem[{Lucarini et~al(2013)Lucarini, Pascale, Boschi, Kirk, and
  Iro}]{ASNA:ASNA201311903}
Lucarini V, Pascale S, Boschi R, Kirk E, Iro N (2013) Habitability and
  multistability in {Earth-like} planets. Astronomische Nachrichten
  334(6):576--588, \doi{10.1002/asna.201311903},
  \urlprefix\url{http://dx.doi.org/10.1002/asna.201311903}

\bibitem[{Madr\'e(2011)}]{Tobias:2011}
Madr\'e TK (2011) Turbulence transition in the asymptotic suction boundary
  layer, diplomarbeit, Philipps-Universit\"at, Marburg, Germany

\bibitem[{Marotzke and Botzet(2007)}]{GRL:GRL22754}
Marotzke J, Botzet M (2007) Present-day and ice-covered equilibrium states in a
  comprehensive climate model. Geophysical Research Letters 34(16):n/a--n/a,
  \doi{10.1029/2006GL028880},
  \urlprefix\url{http://dx.doi.org/10.1029/2006GL028880}

\bibitem[{North et~al(1981)North, Cahalan, and Coakley}]{ROG:ROG799}
North GR, Cahalan RF, Coakley JA (1981) Energy balance climate models. Reviews
  of Geophysics 19(1):91--121, \doi{10.1029/RG019i001p00091},
  \urlprefix\url{http://dx.doi.org/10.1029/RG019i001p00091}

\bibitem[{Nusse and Yorke(1989)}]{Nusse1989137}
Nusse HE, Yorke JA (1989) A procedure for finding numerical trajectories on
  chaotic saddles. Physica D: Nonlinear Phenomena 36(1-2):137 -- 156,
  \doi{http://dx.doi.org/10.1016/0167-2789(89)90253-4},
  \urlprefix\url{http://www.sciencedirect.com/science/article/pii/0167278989902534}

\bibitem[{Paltridge(1978)}]{Paltridge:1978}
Paltridge GW (1978) The steady-state format of global climate. Quarterly
  Journal of the Royal Meteorological Society 104(442):927--945,
  \doi{10.1002/qj.49710444206},
  \urlprefix\url{http://dx.doi.org/10.1002/qj.49710444206}

\bibitem[{Pascale et~al(2012)Pascale, Gregory, Ambaum, Tailleux, and
  Lucarini}]{PGATL:2012}
Pascale S, Gregory JM, Ambaum MHP, Tailleux R, Lucarini V (2012) Vertical and
  horizontal processes in the global atmosphere and the maximum entropy
  production conjecture. Earth System Dynamics 3(1):19--32,
  \doi{10.5194/esd-3-19-2012},
  \urlprefix\url{http://www.earth-syst-dynam.net/3/19/2012/}

\bibitem[{Pierrehumbert(2005)}]{JGRD:JGRD11691}
Pierrehumbert RT (2005) Climate dynamics of a hard snowball earth. Journal of
  Geophysical Research: Atmospheres 110(D1), \doi{10.1029/2004JD005162},
  \urlprefix\url{http://dx.doi.org/10.1029/2004JD005162}

\bibitem[{Pierrehumbert et~al(2011)Pierrehumbert, Abbot, Voigt, and
  Koll}]{ISI:000291366800015}
Pierrehumbert RT, Abbot DS, Voigt A, Koll D (2011) Climate of the
  neoproterozoic. Annual Review of Earth and Planetary Sciences, vol~39, pp
  417--460, \doi{10.1146/annurev-earth-040809-152447}

\bibitem[{Saltzman(2002)}]{Saltzman}
Saltzman B (2002) Dynamical Paleoclimatology: Generalized Theory of Global
  Climate Change. Academic Press

\bibitem[{Schneider and Eckhardt(2009)}]{Schneider13022009}
Schneider TM, Eckhardt B (2009) Edge states intermediate between laminar and
  turbulent dynamics in pipe flow. Phil Trans R Soc A 367(1888):577--587,
  \doi{10.1098/rsta.2008.0216},
  \urlprefix\url{http://rsta.royalsocietypublishing.org/content/367/1888/577.abstract},
  \eprint{http://rsta.royalsocietypublishing.org/content/367/1888/577.full.pdf+html}

\bibitem[{Schneider et~al(2008)Schneider, Gibson, Lagha, De~Lillo, and
  Eckhardt}]{PhysRevE.78.037301}
Schneider TM, Gibson JF, Lagha M, De~Lillo F, Eckhardt B (2008)
  Laminar-turbulent boundary in plane {Couette} flow. Phys Rev E 78:037,301,
  \doi{10.1103/PhysRevE.78.037301},
  \urlprefix\url{http://link.aps.org/doi/10.1103/PhysRevE.78.037301}

\bibitem[{Sellers(1969)}]{Sellers:1969}
Sellers WD (1969) A global climatic model based on the energy balance of the
  earth-atmosphere system. J Appl Meteor 8(3):392--400

\bibitem[{Sieber and Thompson(2012)}]{Sieber13032012}
Sieber J, Thompson JMT (2012) Nonlinear softening as a predictive precursor to
  climate tipping. Phil Trans R Soc A 370(1962):1205--1227,
  \doi{10.1098/rsta.2011.0372},
  \urlprefix\url{http://rsta.royalsocietypublishing.org/content/370/1962/1205.abstract},
  \eprint{http://rsta.royalsocietypublishing.org/content/370/1962/1205.full.pdf+html}

\bibitem[{Skufca et~al(2006)Skufca, Yorke, and
  Eckhardt}]{PhysRevLett.96.174101}
Skufca JD, Yorke JA, Eckhardt B (2006) Edge of chaos in a parallel shear flow.
  Phys Rev Lett 96:174,101, \doi{10.1103/PhysRevLett.96.174101},
  \urlprefix\url{http://link.aps.org/doi/10.1103/PhysRevLett.96.174101}

\bibitem[{Stommel(1961)}]{Stommel:1961}
Stommel H (1961) Thermohaline convection with two stable regimes of flow.
  Tellus 13:224--230

\bibitem[{Stone(1978)}]{Stone:1978}
Stone PH (1978) Baroclinic adjustment. Journal of the Atmospheric Sciences
  35(4):561--571, \doi{10.1175/1520-0469(1978)035<0561:BA>2.0.CO;2},
  \urlprefix\url{http://dx.doi.org/10.1175/1520-0469(1978)035<0561:BA>2.0.CO;2}

\bibitem[{T\'el and Gruiz(2006)}]{TG:2006}
T\'el T, Gruiz M (2006) Chaotic Dynamics. Cambridge University Press, Cambridge

\bibitem[{T\'el et~al(2008)T\'el, Lai, and Gruiz}]{TLG:2008}
T\'el T, Lai YC, Gruiz M (2008) Noise-induced chaos: A consequence of long
  deterministic transients. International Journal of Bifurcation and Chaos
  18(02):509--520, \doi{10.1142/S0218127408020422},
  \urlprefix\url{http://www.worldscientific.com/doi/abs/10.1142/S0218127408020422},
  \eprint{http://www.worldscientific.com/doi/pdf/10.1142/S0218127408020422}

\bibitem[{Voigt and Marotzke(2010)}]{VM:2010}
Voigt A, Marotzke J (2010) {The transition from the present-day climate to a
  modern Snowball Earth}. Climate Dynamics 35(5):887--905,
  \doi{10.1007/s00382-009-0633-5},
  \urlprefix\url{http://dx.doi.org/10.1007/s00382-009-0633-5}

\bibitem[{Wetherald and Manabe(1975)}]{WM:1975}
Wetherald RT, Manabe S (1975) The effect of changing the solar constant on the
  climate of a general circulation model. J Atmos Sci 32:2044--2059

\bibitem[{Zaliapin and Ghil(2010)}]{npg-17-113-2010}
Zaliapin I, Ghil M (2010) Another look at climate sensitivity. Nonlinear
  Processes in Geophysics 17(2):113--122, \doi{10.5194/npg-17-113-2010},
  \urlprefix\url{http://www.nonlin-processes-geophys.net/17/113/2010/}

\end{thebibliography}



\end{document}